\documentclass[letterpaper, 10 pt, conference]{ieeeconf}%
\usepackage{bm}
\usepackage{amssymb}
\usepackage{epsfig}
\usepackage{amsmath}
\usepackage{graphicx}
\usepackage{epstopdf}
\usepackage{amsfonts}%
\usepackage[noend]{algpseudocode}
\usepackage{algorithmicx,algorithm}
\providecommand{\U}[1]{\protect\rule{.1in}{.1in}}
\IEEEoverridecommandlockouts
\begin{document}

\title{{\LARGE \textbf{Decentralized Optimal Merging Control for Connected and
Automated Vehicles }}}
\author{Wei Xiao and Christos G. Cassandras$^{1}$\thanks{*This work was supported in
part by NSF under grants ECCS-1509084, IIP-1430145 and CNS-1645681, by AFOSR
under grant FA9550-12-1-0113, by ARPA-E's NEXTCAR program under grant
DE-AR0000796, and by Bosch and the MathWorks.}\thanks{$^{1}$The authors are
with the Division of Systems Engineering and Center for Information and
Systems Engineering, Boston University, Brookline, MA, 02446, USA
\texttt{{\small \{xiaowei,cgc\}@bu.edu}}}}
\maketitle

\begin{abstract}
This paper addresses the optimal control of Connected and Automated Vehicles
(CAVs) arriving from two roads at a merging point where the objective is to
jointly minimize the travel time and energy consumption of each CAV. The
solution guarantees that a speed-dependent safety constraint is always
satisfied, both at the merging point and everywhere within a control zone
which precedes it. We first analyze the case of no active constraints and
prove that under certain conditions the safety constraint remains inactive,
thus significantly simplifying the determination of an explicit decentralized
solution. When these conditions do not apply, an explicit solution is still
obtained that includes intervals over which the safety constraint is active.
Our analysis allows us to study the tradeoff between the two objective
function components (travel time and energy within the control zone).
Simulation examples are included to compare the performance of the optimal
controller to a baseline with human-driven vehicles with results showing
improvements in both metrics.

\end{abstract}

\thispagestyle{empty} \pagestyle{empty}



\section{INTRODUCTION}

Traffic management at merging points (usually, highway on-ramps) is one of the
most challenging problems within a transportation system in terms of safety,
congestion, and energy consumption, in addition to being a source of stress
for many drivers \cite{Schrank2015, Tideman2007, Waard2009}. Advancements in
next generation transportation system technologies and the emergence of CAVs
(also known as self-driving cars or autonomous vehicles) have the potential to
drastically improve a transportation network's performance by better assisting
drivers in making decisions, ultimately reducing energy consumption, air
pollution, congestion and accidents. One of the very early efforts exploiting
the benefit of CAVs was proposed in \cite{Levine1966}, where an optimal linear
feedback regulator is introduced for the merging problem to control a single
string of vehicles. An overview of automated intelligent vehicle-highway
systems was provided in \cite{Varaiya1993}.

There has been significant research in assisted freeway merging offering
guidance to drivers so as to avoid congestion and collisions. A Classification
and Regression Tree (CART) method was used in \cite{Weng2016} to model merging
behavior and assist decisions in terms of the time-to-collision between
vehicles. The Long Short-Term Memory (LSTM) network was used in
\cite{Chen2017} to predict possible long-term congestion. In \cite{Zang2009},
a Radial Basis Function-Artificial Neural Networks (RBF-ANN) is used to
forcast the traffic volume in a merging area. However, such assisted merging
methods do not take advantage of autonomous driving so as to possibly automate
the merging process in a cooperative manner.

A number of centralized or decentralized merging control mechansims have been
proposed \cite{Ntousakis2016, Cao2015, Milanes2012, Mukai2017, Tideman2007,
Torres2015, Raravi2007, Scarinci2014}. In the case of decentralized control,
all computation is performed on board each vehicle and shared only with a
small number of other vehicles which are affected by it. Optimal control
problem formulations are used in some of these approaches, while Model
Predictive Control (MPC) techniques are employed in others, primarily to
account for additional constraints and to compensate for disturbances by
re-evaluating optimal actions. The objectives specified for optimal control
problems may target the minimization of acceleration as in \cite{Torres2015}
or the maximization of passenger comfort (measured as the acceleration
derivative or jerk) as in \cite{Ntousakis2016, Rathgeber2015}. MPC approaches
have been used in \cite{Cao2015, Mukai2017}, as well as in
\cite{Ntousakis2016} when inequality constraints are added to the originally
considered optimal control problem.

In \cite{Zhang2016}, a decentralized optimal control framework is provided for
a signal-free intersection. This may be viewed as a process of merging
multiple traffic flows so that the highway merging problem is a special case.
However, as detailed in the sequel, there are several differences in the
formulation and analysis we pursue here in terms of the objective function and
the safety constraints used.

In this paper, we develop a decentralized optimal control framework for each
CAV approaching a merging point from one of two roads (often, a highway lane
and an on-ramp lane). Our objective differs from formulations in
\cite{Ntousakis2016}, \cite{Torres2015} or \cite{Zhang2016}; moreover, it is
designed to guarantee that a hard speed-dependent safety constraint is always
satisfied. In particular, our objective combines minimizing $(i)$ the travel
time of each CAV over a given road segment from a point entering a Control
Zone (CZ) to the eventual Merging Point (MP) and $(ii)$ a measure of its
energy consumption. This allows us to explore the trade-off between these two
metrics as a function of a weight factor. The problem incorporates CAV speed
and acceleration constraints, and a hard safety constraint requiring a minimal
headway between adjacent vehicles at all times as well as guaranteed collision
avoidance at the MP. We derive an analytical solution of the problem and
identify several properties of an optimal trajectory. This allows us to obtain
simple to check conditions under which the safe distance constraint is
guaranteed to not become active (which significantly reduces computation); in
cases where it does become active, we include constrained arcs as part of an
optimal trajectory . Thus, we can identify when a trajectory exists that
provably satisfies all constraints at all times and explicitly determine the
optimal merging trajectory of each CAV.

The paper is structured as follows. In Section II, we present the merging
process model and formulate the optimal merging control problem including all
safety requirements that must be satisfies at all times. In Section III, the
optimal solutions in all cases are presented. We show the simulations and
discussion in Section IV and V, respectively.

\section{PROBLEM FORMULATION}

\begin{figure}[ptbh]
\centering
\includegraphics[scale=0.22]{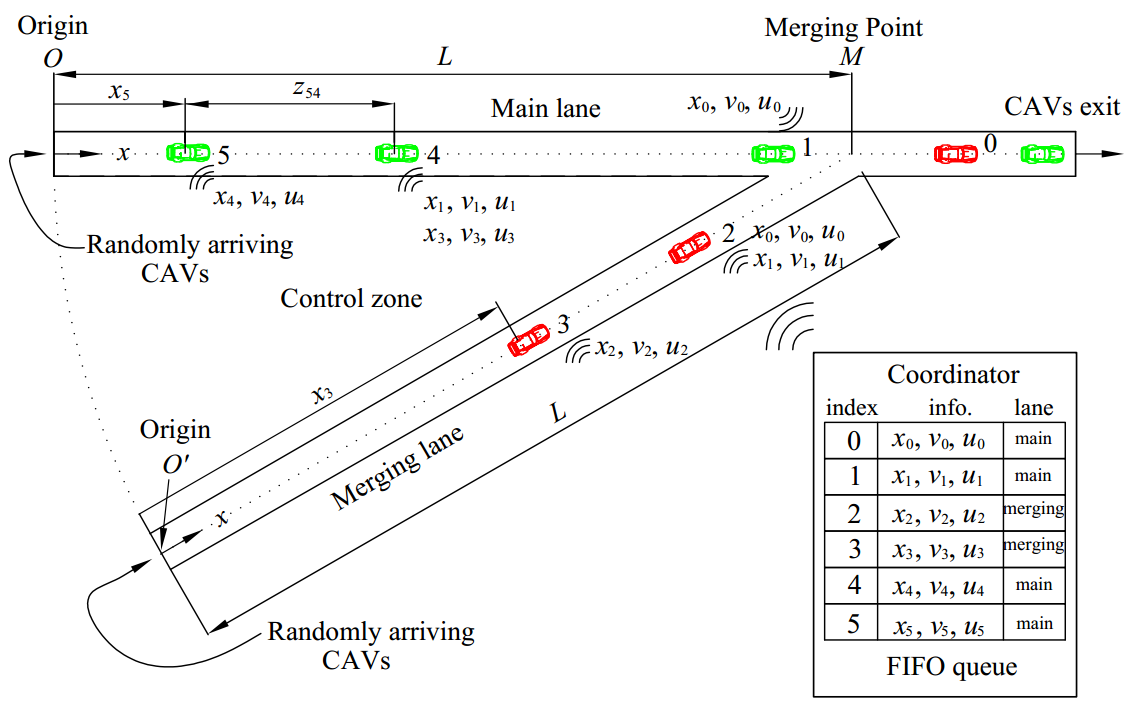} \caption{The merging problem}%
\label{modelF}%
\end{figure}

The merging problem arises when traffic must be joined from two different
roads, usually associated with a main lane and a merging lane as shown in
Fig.\ref{modelF}. We consider the case where all traffic consists of CAVs randomly
arriving at the two lanes joined at the Merging Point (MP) $M$ where a
collison may occur. The segment from the origin $O$ or $O^{\prime}$ to the
merging point $M$ has a length $L$ for both lanes, and is called the Control
Zone (CZ). We assume that CAVs do not overtake each other in the CZ. A
coordinator is associated with the MP whose function is to maintain a
First-In-First-Out (FIFO) queue of CAVs based on their arrival time at the CZ
and enable real-time communication with the CAVs that are in the CZ as well as
the last one leaving the CZ. The FIFO assumption imposed so that CAVs cross
the MP in their order of arrival is made for simplicity and often to ensure
fairness, but can be relaxed through dynamic resequencing schemes, e.g., as
described in \cite{Zhang2018}.

Let $S(t)$ be the set of FIFO-ordered indices of all CAVs located in the CZ at
time $t$ along with the CAV (whose index is 0 as shown in Fig.\ref{modelF}) that has just
left the CZ. Let $N(t)$ be the cardinality of $S(t)$. Thus, if a CAV arrives
at time $t$ it is assigned the index $N(t)$. All CAV indices in $S(t)$
decrease by one when a CAV passes over the MP and the vehicle whose index is
$-1$ is dropped.

The vehicle dynamics for each CAV $i\in S(t)$ along the lane to which it
belongs take the form
\begin{equation}
\left[
\begin{array}
[c]{c}%
\dot{x}_{i}(t)\\
\dot{v}_{i}(t)
\end{array}
\right]  =\left[
\begin{array}
[c]{c}%
v_{i}(t)\\
u_{i}(t)
\end{array}
\right]  \label{VehicleDynamics}%
\end{equation}
where $x_{i}(t)$ denotes the distance to the origin $O$ ($O^{\prime}$) along
the main (merging) lane if the vehicle $i$ is located in the main (merging)
lane, $v_{i}(t)$ denotes the velocity, and $u_{i}(t)$ denotes the control
input (acceleration). We consider two objectives for each CAV subject to three
constraints, as detailed next.

$\mathbf{Objective1}$ (Minimizing travel time): Let $t_{i}^{0}$ and $t_{i}%
^{m}$ denote the time that CAV $i\in S(t)$ arrives at the origin $O$ or
$O^{\prime}$ and the merging point $M$, respectively. We wish to minimize the
travel time $t_{i}^{m}-t_{i}^{0}$ for CAV $i$.

$\mathbf{Objective2}$ (Minimizing energy consumption): We also wish to
minimize energy consumption for each CAV $i\in S(t)$ expressed as
\begin{equation}
J_{i}(t_{i}^{m},u_{i}(t))=\int_{t_{i}^{0}}^{t_{i}^{m}}C(u_{i}(t))dt,
\end{equation}
where $C(\cdot)$ is a strictly increasing function of its argument.

$\mathbf{Constraint1}$ (Safety constraints): Let $i_{p}$ denote the index of
the CAV which physically immediately precedes $i$ in the CZ (if one is
present). We require that the distance $z_{i,i_{p}}(t):=x_{i_{p}}(t)-x_{i}(t)$
be constrained by the speed $v_{i}(t)$ of CAV $i\in S(t)$ so that
\begin{equation}
z_{i,i_{p}}(t)\geq\varphi v_{i}(t)+\delta,\text{ \ }\forall t\in\lbrack
t_{i}^{0},t_{i}^{m}], \label{Safety}%
\end{equation}
where $\varphi$ denotes the reaction time (as a rule, $\varphi=1.8$ is used,
e.g., \cite{Vogel2003}). If we define $z_{i,i_{p}}$ to be the distance from
the center of CAV $i$ to the center of CAV $i_{p}$, then $\delta$ is a
constant determined by the length of these two CAVs (generally dependent on
$i$ and $i_{p}$ but taken to be a constant over all CAVs for simplicity).

$\mathbf{Constraint2}$ (Safe merging): There should be enough safe space at
the MP $M$ for a merging CAV to cut in, i.e.,
\begin{equation}
z_{1,0}(t_{1}^{m})\geq\varphi v_{1}(t_{1}^{m})+\delta. \label{SafeMerging}%
\end{equation}

$\mathbf{Constraint3}$ (Vehicle limitations): Finally, there are constraints
on the speed and acceleration for each $i\in S(t)$, i.e.,
\begin{equation}
\begin{aligned} v_{min} \leq v_i(t)\leq v_{max}, \forall t\in[t_i^0,t_i^m],\\ u_{min}\leq u_i(t)\leq u_{max}, \forall t\in[t_i^0,t_i^m], \end{aligned} \label{VehicleConstraints}%
\end{equation}
where $v_{max}>0$ and $v_{min}>0$ denote the maximum and minimum speed allowed
in the CZ, while $u_{min}<0$ and $u_{max}>0$ denote the minimum and maximum
control input, respectively.

\textbf{Problem Formulation. }Our goal is to determine a control law to
achieve objectives 1-2 subject to constraints 1-3 for each $i\in S(t)$
governed by the dynamics (\ref{VehicleDynamics}). Combining objectives 1 and
2, we formulate the following optimal control problem for each CAV:
\begin{equation}
\min_{u_{i}(t)}J_{i}(t_{i}^{m},u_{i}(t)):=\beta(t_{i}^{m}-t_{i}^{0}%
)+\int_{t_{i}^{0}}^{t_{i}^{m}}\frac{1}{2}u_{i}^{2}(t)dt, \label{CAV_problem}%
\end{equation}
subject to (\ref{VehicleDynamics}), (\ref{Safety}), (\ref{SafeMerging}),
(\ref{VehicleConstraints}), the initial and terminal position conditions
$x_{i}(t_{i}^{0})=0$, $x_{i}(t_{i}^{m})=L$, and given $t_{i}^{0},v_{i}%
(t_{i}^{0})$. The weight factor $\beta\geq0$ can be adjusted to penalize
travel time relative to the energy cost. The two terms in (\ref{CAV_problem})
need to be properly normalized. Thus, by defining $t_{i,max}=L/v_{min}$ to be
the maximum travel time and using $\alpha\in\lbrack0,1]$, we construct a
convex combination as follows:
\begin{equation}
\begin{aligned} J_i(t_i^m, u_i(t))= &\alpha\frac{(t_i^m - t_i^0)}{t_{i,max}} + \frac{(1-\alpha)\int_{t_i^0}^{t_i^m}\frac{1}{2}u_i^2(t)dt}{\frac{1}{2}\max(u_{max}^2,u_{min}^2)t_{i,max}}\\ = &\frac{\alpha \max(u_{max}^2,u_{min}^2)}{2(1-\alpha)} (t_i^m\! -\! t_i^0)\! +\! \int_{t_i^0}^{t_i^m}\!\frac{1}{2}u_i^2(t)dt \end{aligned}
\end{equation}
We can then set $\beta=\frac{\alpha\max(u_{max}^{2},u_{min}^{2})}{2(1-\alpha
)}$ and use (\ref{CAV_problem}) as the problem to be solved.

\section{DECENTRALIZED FRAMEWORK}

Note that (\ref{CAV_problem}) can be locally solved by each CAV $i$ provided
that there is some information sharing with two other CAVs: CAV $i_{p}$ which
physically immediately precedes $i$ and is needed in (\ref{Safety}) and CAV
$i-1$ so that $i$ can determine whether this CAV is located in the same lane
or not. With this information, CAV $i$ can determine which of two possible
cases applies: $(i)$ $i_{p}=i-1$, i.e., $i_{p}$ is the CAV immediately
preceding $i$ in the FIFO queue (e.g., CAVs 3 and 5 in Fig.\ref{modelF}), and $(ii)$
$i_{p}<i-1$, which implies\ that CAV $i-1$ is in a different lane from $i$
(e.g., CAVs 2 and 4 in Fig.\ref{modelF}). It is now clear that we can solve problem
(\ref{CAV_problem}) for any $i\in S(t)$ in a decentralized way in the sense
that CAV $i$ needs only its own local information and information from $i-1$,
as well as from $i_{p}$ in case $(ii)$. Observe that if $i_{p}=i-1$, then
(\ref{SafeMerging}) is a redundant constraint; otherwise, we need to
separately consider (\ref{Safety}) and (\ref{SafeMerging}). Therefore, we will
analyze each of these two cases in what follows.

\subsection{Decentralized Optimal Control when $i-1=i_{p}$}

Let $\bm x_{i}(t):=(x_{i}(t),v_{i}(t))^T$ be the state
vector and $\bm\lambda_{i}(t):=(\lambda_{i}^{x}(t),\lambda_{i}^{v}%
(t))^T$ be the costate vector (for
simplicity, in the sequel we omit explicit time dependence when no ambiguity
arises). The Hamiltonian with the state constraint, control constraint and
safety constraint adjoined is
\begin{equation}
\begin{aligned} H_i(\bm x_i,\bm\lambda_i, u_i) = &\frac{1}{2}u_i^2\! +\! \lambda_i^xv_i+ \lambda_i^vu_i \\&+ \mu_i^a(u_i\! -\! u_{max}) + \mu_i^b(u_{min} - u_i) \\&+ \mu_i^c(v_i - v_{max}) + \mu_i^d(v_{min} - v_i) \\&+ \mu_i^e(x_i + \varphi v_i + \delta -x_{i_p}) + \beta \end{aligned}
\end{equation}
The Lagrange multipliers $\mu_{i}^{a},\mu_{i}^{b},\mu_{i}^{c},\mu_{i}^{d},\mu_{i}^{e}$ are positive when the constraints are active and become 0 when the constraints are strict. Note that when the safety constraint (\ref{Safety}) becomes active, the expression above involves $x_{i_p}(t)$ in the last term. When $i=1$, the optimal trajectory is obtained without this term, since (\ref{Safety}) is inactive over all $[t_1^0,t_1^m]$. Thus, once the solution for $i=1$ is obtained (based on the analysis that follows), $x_1^{\ast}$ is a given function of time and available to $i=2$. Based on this information, the optimal trajectory of $i=2$ is obtained. Similarly, all subsequent optimal trajectories for $i >2$ can be recursively obtained based on $x_{i_p}^{\ast}(t)$ with $i_p = i-1$.

Since $\psi_{i,1}:=x_{i}%
(t_{i}^{m})-L=0$ is not an explicit function of time, the transversality
condition \cite{Bryson1969} is
\begin{equation}
\left.  H_{i}(\bm x_{i}(t),\bm\lambda_{i}(t),u_{i}(t))\right\vert
_{t=t_{i}^{m}}=0 \label{Transversality}%
\end{equation}
with the costate boundary condition $\bm\lambda_{i}(t_{i}^{m})=[(\nu
_{i,1}\frac{\partial\psi_{i,1}}{\partial\bm x_{i}})^{T}]_{t=t_{i}^{m}}$, where
$\nu_{i,1}$ denotes a Lagrange multiplier.

The Euler-Lagrange equations become
\begin{equation}
\dot{\lambda}_{i}^{x}=-\frac{\partial H_{i}}{\partial x_{i}}=-\mu_{i}^{e}%
 \label{EulerEqX}%
\end{equation}
and
\begin{equation}
 \dot \lambda_i^v = -\frac{\partial H_i}{\partial v_i} = -\lambda_i^x - \mu_i^c + \mu_i^d - \varphi\mu_i^e,  \label{EulerEqV}%
\end{equation}
and the necessary condition for optimality is
\begin{equation}
\frac{\partial H_{i}}{\partial u_{i}}=u_{i}+\lambda_{i}^{v}+\mu_{i}^{a}%
-\mu_{i}^{b}=0. \label{OptimalityCondition}%
\end{equation}

\vspace{2ex} $\mathbf{Assumption1:}$ The safety constraint (\ref{Safety}),
control and state constraints (\ref{VehicleConstraints}) are not active at
$t_{i}^{0}$.

Since CAVs arrive randomly, there are two ways to handle violations of
Assumption 1: $(i)$ By ensuring that it holds through a Feasibility
Enforcement Zone (FEZ) as in \cite{Zhang2017} which applies the necessary
control prior to the CZ so as to enforce (\ref{Safety}) and
(\ref{VehicleConstraints}) upon arrival at the CZ, $(ii)$ by foregoing
optimality and simply controlling a CAV that violates Assumption 1 until all
constraints become feasible within the CZ.

Under Assumption 1, we will start by analyzing the case of no active
constraints and then study what happens as different constraints become
active. In this paper, we limit ourselves to cases where (\ref{Safety}) may
become active which are much more challenging than (\ref{VehicleConstraints});
the latter can also be handled through an analysis similar to that found in
\cite{Malikopoulos2018}.

\subsubsection{\textbf{Control, state, safety constraints not active}}
\label{sec:unconstrained_A}

In this case, $\mu_{i}^{a}=\mu_{i}^{b}=\mu_{i}^{c}=\mu_{i}^{d}=\mu_{i}^{e}=0$.
Applying (\ref{OptimalityCondition}), the optimal control input is given by
\begin{equation}
u_{i}+\lambda_{i}^{v}=0. \label{OptimalUinLambda}%
\end{equation}
and the Euler-Lagrange equation (\ref{EulerEqV}) yields
\begin{equation}
\dot{\lambda}_{i}^{v}=-\lambda_{i}^{x}.
\end{equation}
Therefore, (\ref{EulerEqX}) implies $\lambda_{i}^{x}(t)=a_{i}$, hence
$\lambda_{i}^{v}(t)=-(a_{i}t+b_{i})$, where $a_{i}$ and $b_{i}$ are
integration constants. Consequently, we obtain the following optimal
solution:
\begin{equation}
u_{i}^{\ast}(t)=a_{i}t+b_{i} \label{Optimal_u}%
\end{equation}%
\begin{equation}
v_{i}^{\ast}(t)=\frac{1}{2}a_{i}t^{2}+b_{i}t+c_{i} \label{Optimal_v}%
\end{equation}%
\begin{equation}
x_{i}^{\ast}(t)=\frac{1}{6}a_{i}t^{3}+\frac{1}{2}b_{i}t^{2}+c_{i}t+d_{i}
\label{Optimal_x}%
\end{equation}
where $c_{i}$ and $d_{i}$ are also integration constants. In addition, we have
the initial conditions $x_{i}(t_{i}^{0})=0,v_{i}(t_{i}^{0})=v_{i}^{0}$ and the
terminal condition $x_{i}(t_{i}^{m})=L$. The costate boundary conditions and
(\ref{OptimalityCondition}) offer us $u_{i}(t_{i}^{m})=-\lambda_{i}^{v}%
(t_{i}^{m})=0$ and $\bm\lambda_{i}(t_{i}^{m})=(a_{i},0)$, therefore, the
transversality condition (\ref{Transversality}) gives us an additional
relationship:
\begin{equation}
\beta+a_{i}v_{i}(t_{i}^{m})=0. \label{TransEqInA}%
\end{equation}
Then, for each $i\in S(t)$, we need to solve the following five nonlinear
algebraic equations for $a_{i},b_{i},c_{i},d_{i}$ and $t_{i}^{m}$:
\begin{equation}
\begin{aligned} &\frac{1}{2}a_i\cdot(t_i^0)^2 + b_it_i^0 + c_i = v_i^0,\\ &\frac{1}{6}a_i\cdot(t_i^0)^3 + \frac{1}{2}b_i\cdot(t_i^0)^2 + c_it_i^0+d_i = 0,\\ &\frac{1}{6}a_i\cdot(t_i^m)^3 + \frac{1}{2}b_i\cdot(t_i^m)^2 + c_it_i^m+d_i = L,\\ &a_it_i^m + b_i = 0,\\ &\beta + \frac{1}{2}a_i^2\cdot(t_i^m)^2 + a_ib_it_i^m + a_ic_i = 0. \end{aligned} \label{OptimalSolInA}%
\end{equation}

There may be four, six or eight solutions if we solve (\ref{OptimalSolInA}),
depending on the values of $t_{i}^{0},\beta,L$ and $v_{i}^{0}$, but only one
of the solutions is valid, i.e., it satisfies $t_{i}^{m}>t_{i}^{0}$ and
$t_{i}^{m}$ is a real number. The remaining solutions are either imaginary or
negative numbers. The following six lemmas provide a number of useful
properties of the optimal solution (\ref{Optimal_u})-(\ref{Optimal_x}). 

Observe that when $\beta = 0$, it follows that $a_i = 0$ from (19). Then, we can easily get the obvious solution 
\begin{equation}
t_i^m - t_i^0 = \frac{L}{v_i^0}. \label{OptimalTimeBelta0}%
\end{equation}

\vspace{2ex}\textbf{Lemma 1:} The optimal terminal time $t_{i}^{m}$ can be
expressed as a polynomial equation in the known parameters $t_{i}^{0}$,
$\beta,L$ and $v_{i}^{0}$.\vspace{2ex}

$\emph{Proof:}$ If $\beta = 0$, the result is true from (\ref{OptimalTimeBelta0}). If $\beta > 0$, then combining the first and second equations of (\ref{OptimalSolInA}), we get
\begin{equation}
\frac{1}{3}b_i(t_i^0)^2 + (\frac{2}{3}v_i^0 + \frac{4}{3}c_i)t_i^0 + 2d_i = 0.
\end{equation}
Combining the third and fourth equations of (\ref{OptimalSolInA}), we get
\begin{equation}
\frac{1}{3}b_i(t_i^m)^2 + c_it_i^m + d_i = L. \label{Lemma1Eq1}%
\end{equation}
Combining the last two equations, we get
\begin{equation}
\frac{1}{3}b_i((t_i^m)^2 - (t_i^0)^2) + (\frac{2}{3}v_i^0 + \frac{4}{3}c_i)(t_i^m - t_i^0) + \frac{b_ic_i}{3a_i} +\frac{2b_iv_i^0}{3a_i}  = L + d_i.
\end{equation}
Subtracting the first equation from the last equation of (\ref{OptimalSolInA}),
\begin{equation}
\frac{1}{2}a_i((t_i^m)^2 - (t_i^0)^2) + b_i(t_i^m - t_i^0)   = -\frac{\beta}{a_i} - v_i^0. \label{LemmaEq4}
\end{equation}
Then, combining the last two equations, we get
\begin{equation}
t_i^m - t_i^0  = \frac{\frac{a_i}{2}(L+d_i) - \frac{b_ic_i}{6}+ \frac{b_i\beta}{3a_i}}{\frac{a_iv_i^0}{3} + \frac{2a_ic_i}{3} - \frac{b_i^2}{3}}. \label{TimeComplex}
\end{equation}
Combining (\ref{Lemma1Eq1}) and the last two equations of (\ref{OptimalSolInA}), we get
\begin{equation}
-\frac{2b_i^3}{3a_i^2} + \frac{5b_ic_i}{3a_i}+ \frac{2b_i\beta}{3a_i^2} = d_i - L
\end{equation}
Taking the square of the fourth equation of (\ref{OptimalSolInA}) and combining with the last equation of (\ref{OptimalSolInA}) yields
\begin{equation}
b_i^2 = 2\beta + 2a_ic_i \label{Lemma1Eq2}
\end{equation}
Combining the last two equations, we get
\begin{equation}
\frac{a_id_i}{2} = \frac{a_iL}{2} - \frac{b_i\beta}{3a_i} + \frac{b_ic_i}{6} \label{Lemma1Eq3}
\end{equation}
Combining (\ref{Lemma1Eq3}) and (\ref{Lemma1Eq2}) with the numerator and denominator of (\ref{TimeComplex}), respectively, we get
\begin{equation}
t_i^m - t_i^0  = \frac{3a_iL}{a_iv_i^0 - 2\beta} \label{TravelTime}
\end{equation}
Combining (\ref{LemmaEq4}) and (\ref{TravelTime}), we get
\begin{equation}
(t_i^m)^2 - (t_i^0)^2  = -\frac{2\beta}{a_i^2} - \frac{2v_i^0}{a_i} - \frac{6b_iL}{a_iv_i^0 - 2\beta} \label{LemmaEq5}
\end{equation}
Subtracting the second equation from the third equation of (\ref{OptimalSolInA}), we get
\begin{equation}
\frac{1}{6}a_i((t_i^m)^3 - (t_i^0)^3) + \frac{1}{2}b_i((t_i^m)^2 - (t_i^0)^2) + c_i(t_i^m- t_i^0) = L\label{LemmaEq6}\end{equation} 
Combining (\ref{Lemma1Eq2}), (\ref{TravelTime}), (\ref{LemmaEq5}) and (\ref{LemmaEq6}) gives
\begin{equation}
\frac{a_i}{6}((t_i^m)^3 - (t_i^0)^3)  -\frac{b_i\beta}{a_i^2} - \frac{b_iv_i^0}{a_i} - \frac{3b_i^2L + 6\beta L}{2(a_iv_i^0 - 2\beta)} = L\label{LemmaEq7}
\end{equation}

Rewriting (\ref{TravelTime}) as
\begin{equation}
a_i = \frac{2\beta(t_i^m - t_i^0)}{(t_i^m - t_i^0)v_i^0 - 3L},\label{LemmaEq8}
\end{equation}
we notice that $a_i$ only depends on $t_i^0, t_i^m, v_i^0, L,\beta$. Rewriting (\ref{LemmaEq5}) as
\begin{equation}
b_i = -\left[\frac{(t_i^m)^2 - (t_i^0)^2}{6L} + \frac{\beta + a_iv_i^0}{3a_i^2L}\right](a_iv_i^0 - 2\beta),\label{LemmaEq9}
\end{equation}
we notice $b_i$ only depends on $t_i^0, t_i^m, v_i^0, L,\beta, a_i$, because $a_i$ only depends on $t_i^0, t_i^m, v_i^0, L,\beta$. Therefore, $b_i$ only depends on $t_i^0, t_i^m, v_i^0, L,\beta$. In (\ref{LemmaEq7}), $t_i^m$ only depends on $t_i^0, v_i^0, L,\beta, a_i, b_i$. So when solving (\ref{LemmaEq7}) for $t_i^m$ with (\ref{LemmaEq8}) and (\ref{LemmaEq9}), the solutions only depend on $t_i^0, v_i^0, L,\beta$.  $\;\;\;\qquad\blacksquare$

\vspace{2ex} \textbf{Lemma 2:} The solution for $a_{i}$ in (\ref{OptimalSolInA}) is independent
of $t_{i}^{0}$. Moreover, $a_{i}\leq0$.\vspace{2ex}

$\emph{Proof:}$ If $\beta = 0$, then $a_i = 0$ follows from the last equation of (\ref{OptimalSolInA}). Otherwise, combining (\ref{Lemma1Eq2}) and the first equation of (\ref{OptimalSolInA}), we get
\begin{equation}
a_i^2(t_i^0)^2 + 2a_ib_it_i^0 + b_i^2 - 2\beta = 2a_iv_i^0.\label{LemmaEq10}
\end{equation}
Subtracting (\ref{LemmaEq10}) from the square of the fourth equation of (\ref{OptimalSolInA}), we get
\begin{equation}
a_i^2((t_i^m)^2-(t_i^0)^2) + 2a_ib_i(t_i^m-t_i^0) + 2\beta = -2a_iv_i^0.\label{LemmaEq11}
\end{equation}
Combining the fourth equation of (\ref{OptimalSolInA}), (\ref{TravelTime}) and (\ref{LemmaEq11}), we get
\begin{equation}
-a_i^2(\frac{3a_iL}{a_iv_i^0 - \beta})^2 + 2\beta = -2a_iv_i^0.\label{Lemmaa_i}
\end{equation}
where $L, \beta, v_i^0$ are known parameters, and $t_i^0$ does not appear in (\ref{Lemmaa_i}). Therefore, $a_i$ is independent of $t_i^0$. 

In (\ref{TransEqInA}), i.e., $\beta+a_iv_i(t_i^m) = 0$, since $\beta > 0$ and $v_i(t_i^m) > 0$, then $a_i < 0$.$\qquad\qquad\qquad\qquad\qquad\qquad\qquad\qquad\quad\;\;\blacksquare$

\vspace{2ex} \textbf{Lemma 3:} Given $\beta$, $L$ and under optimal control
(\ref{Optimal_u}), if $v_{i}^{0}=v_{j}^{0}$, then $t_{i}^{m}-t_{j}^{m}%
=t_{i}^{0}-t_{j}^{0}$.\vspace{2ex}

$\emph{Proof:}$ If $\beta = 0$, the result is true from (\ref{OptimalTimeBelta0}). Otherwise, by Lemma 2, $a_i = a_j$ in (\ref{TravelTime}), and $L,\beta$ are known. Since $v_i^0 = v_j^0$, it follows that $t_i^m - t_j^m = t_i^0 - t_j^0$. $\qquad\qquad\qquad\quad\;\;\;\blacksquare$

\vspace{2ex} \textbf{Lemma 4:} Under optimal control (\ref{Optimal_u}),
$v_{i}(t_{i}^{m})=-\frac{\beta}{a_{i}}$ for all $i\in S(t)$, and $v_{i}(t)$ is
strictly increasing for all $t\in\lbrack t_{i}^{0},t_{i}^{m}]$ taking its
maximum value at $t=t_{i}^{m}$ when $\beta>0$. Moreover, $\lim\limits_{\beta
\rightarrow0}\frac{\beta}{-a_{i}}=v_{i}^{0}$ and $\lim\limits_{\beta
\rightarrow0}\frac{3a_{i}L}{a_{i}v_{i}^{0}-2\beta}=\lim\limits_{\beta
\rightarrow0}\frac{3L}{v_{i}^{0}+2v_{i}(t_{i}^{m})}=\frac{L}{v_{i}^{0}}$.\vspace{2ex}

$\emph{Proof:}$ We know $u_i(t_i^m) = 0$ from (\ref{Optimal_u}) and the fourth equation of (\ref{OptimalSolInA}). By Lemma 2, if $\beta > 0$, we have $a_i < 0$, therefore, (\ref{Optimal_u}) implies $u_i(t)>0$ for all $t\in[t_i^0,t_i^m)$, hence $v_i(t)$ is strictly increasing for all $t\in [t_i^0, t_i^m]$ and takes its maximum value at $t = t_i^m$. From (\ref{TransEqInA}), we know $v_i(t_i^m) = -\frac{\beta}{a_i}$. Since $v_i(t)$ is strictly increasing for all $t\in [t_i^0, t_i^m]$ when $\beta \ne 0$, we have $v_i(t_i^m) > v_i(t_i^0)$. From (\ref{LemmaEq8}), we can get $\lim\limits_{\beta\rightarrow 0}a_i = 0$, and further $\lim\limits_{\beta\rightarrow 0}b_i = 0$ from the fourth equation of (\ref{OptimalSolInA}). Finally, we can get $\lim\limits_{\beta\rightarrow 0}u_i(t) = 0$ from (\ref{Optimal_u}), thus, $\lim\limits_{\beta\rightarrow 0}\frac{\beta}{-a_i} =\lim\limits_{\beta\rightarrow 0}v_i(t_i^m) = v_i^0$ and $\lim\limits_{\beta\rightarrow 0}\frac{3a_iL}{a_iv_i^0 - 2\beta}  = \frac{L}{v_i^0}$. $\qquad\qquad\qquad\qquad\qquad\qquad\qquad\qquad\qquad\qquad\quad\;\;\;\;\blacksquare$

\vspace{2ex} \textbf{Lemma 5:} Under optimal control (\ref{Optimal_u}), the
travel time for $i\in S(t)$ satisfies $t_{i}^{m}-t_{i}^{0}\leq\frac{L}%
{v_{i}^{0}}$.\vspace{2ex}

$\emph{Proof:}$ If $\beta = 0$, then $t_i^m - t_i^0 = \frac{L}{v_i^0}$ from (\ref{OptimalTimeBelta0}). Otherwise, by Lemma 4, we know $\lim\limits_{\beta\rightarrow 0}t_i^m - t_i^0 = \frac{L}{v_i^0}$. Because $\beta$ is the penalty of $t_i^m - t_i^0$ in (\ref{CAV_problem}), if $\beta$ increases, then $t_i^m - t_i^0$ must decrease or stay the same. Therefore, $t_i^m - t_i^0 \leq \frac{L}{v_i^0}$. $\qquad\blacksquare$

\vspace{2ex} \textbf{Lemma 6: }For two vehicles $i,j\in S(t)$ under optimal
control (\ref{Optimal_u}), if $v_{i}^{0}<v_{j}^{0}$ and $\beta>0$, then
$v_{i}(t_{i}^{m})<v_{j}(t_{j}^{m})$, $t_{i}^{m}-t_{i}^{0}>t_{j}^{m}-t_{j}^{0}$
and $a_{i}<a_{j}<0$. \vspace{2ex}

$\emph{Proof:}$ We rewrite (\ref{Lemmaa_i}) as
\begin{equation}
9a_i^4L^2 = 2a_i^3(v_i^0)^3 - 6a_i^2(v_i^0)^2\beta + 8\beta^3.
\end{equation}
By Lemma 4, we know $v_i(t_i^m) = -\frac{\beta}{a_i}$, and the equality above becomes
\begin{equation}
\frac{9}{2}\beta L^2 = 4(v_i(t_i^m))^4 - 3(v_i(t_i^m))^2(v_i^0)^2 - v_i(t_i^m)(v_i^0)^3 \label{VimVi0S}
\end{equation}
which can be rewritten as
\begin{equation}
\begin{aligned}
\frac{9}{2}\beta L^2 = 3(v_i(t_i^m))^2((v_i(t_i^m))^2 - (v_i^0)^2) \\+ v_i(t_i^m)((v_i(t_i^m))^3 - (v_i^0)^3).\label{VimVi0}
\end{aligned}
\end{equation}
By Lemma 4, $v_i(t_i^m) > v_i^0$, Therefore, if $v_i^0$ decreases, $v_i(t_i^m)$ must decrease in order to satisfy (\ref{VimVi0}) whose left hand side is fixed. Formally, by taking the derivative with respect to $v_i^0$ in (\ref{VimVi0}), we get
\begin{equation}
\frac{\partial v_i(t_i^m)}{\partial v_i^0} = \frac{6(v_i(t_i^m))^2v_i^0 + 3v_i(t_i^m)(v_i^0)^2}{16(v_i(t_i^m))^3 - 6v_i(t_i^m)(v_i^0)^2 - (v_i^0)^3}\label{DeVimVi0}
\end{equation}

By Lemma 4 and (\ref{VehicleConstraints}), $v_i(t_i^m) > v_i^0 > 0$, therefore, both the denominator and numerator of (\ref{DeVimVi0}) are positive, hence $\frac{\partial v_i(t_i^m)}{\partial v_i^0} >0$. Since $v_i(t_i^m)$ is a strictly increasing function with respect to $v_i^0$, if $v_i^0 < v_j^0$, it follows that $v_i(t_i^m) < v_j(t_j^m)$. Further by (\ref{TransEqInA}) and (\ref{TravelTime}), $t_i^m - t_i^0 =\frac{3L}{v_i^0 + 2v_i(t_i^m)}$, therefore, $t_i^m - t_i^0 > t_j^m - t_j^0$. By Lemma 2, $a_i<0$ and $a_j<0$. Since $v_i(t_i^m) = \frac{\beta}{-a_i}$, it follows that $a_i<a_j <0$.$\;\blacksquare$

Using Lemmas 1-6, we can establish Theorem 1 identifying conditions such that
the safety constraint (\ref{Safety}) is never violated for all $t\in\lbrack
t_{i}^{0},t_{i}^{m}]$ in an optimal trajectory. The following assumption
requires that if two CAVs arrive too close to each other, then the first one
maintains its optimal terminal speed past the MP until the second one crosses
it as well. This is to ensure that the first vehicle does not suddenly decelerate and cause the safety constraint to be violated during the last segment of the first vehicle's optimal trajectory. 

\vspace{2ex} \textbf{Assumption 2: }For a given constant $\zeta>\varphi$, any
CAV $i-1\in S(t)$ such that $t_{i}^{0}-t_{i-1}^{0}<\zeta$ maintains a constant
speed $v_{i-1}(t)=v_{i-1}^{\ast}(t_{i-1}^{m})$ for all $t\in\lbrack
t_{i-1}^{m},t_{i}^{m}]$.

\vspace{2ex} \textbf{Theorem 1: }Under Assumptions 1-2, if CAVs $i$ and $i_{p}$
satisfy $v_{i}^{0}\leq v_{i_{p}}^{0}$ and $t_{i}^{0}-t_{ip}^{0}\geq
\varphi+\frac{\delta}{v_{i}^{0}}$, then, under optimal control
(\ref{Optimal_u}), $z_{i,i_{p}}(t)\geq\varphi v_{i}(t)+\delta$ for all
$t\in\lbrack t_{i}^{0},t_{i}^{m}]$. Moreover, if $\beta>0$, then $z_{i,i_{p}%
}(t)>\varphi v_{i}(t)+\delta$ for all $t\in\lbrack t_{i}^{0},t_{i}^{m}%
]$.\vspace{2ex}

$\emph{Proof:}$ If $\beta=0$, it follows from (\ref{TransEqInA}) that
$a_{i}=a_{i_{p}}=0$, and by the costate boundary conditions, we have $b_i = 0$. Therefore, it follows from (\ref{Optimal_u}) that $u_{i}(t)=u_{i_{p}}(t)=0$, which impies $v_{i}(t)=v_{i}^{0}$
and $v_{i_{p}}(t)=v_{i_{p}}^{0}$ for all $t\in\lbrack t_{i}^{0},t_{i}^{m}]$.
Because $t_{i}^{0}-t_{ip}^{0}\geq\varphi+\frac{\delta}{v_{i}^{0}}$ and
$v_{i}^{0}\leq v_{i_{p}}^{0}$, it follows that $z_{i,i_{p}}(t)\geq\varphi
v_{i}(t)+\delta$ for all $t\in\lbrack t_{i}^{0},t_{i}^{m}]$.

If $\beta>0$, let us first consider the case $v_{i}^{0}=v_{i_{p}}^{0}$. Since
$t_{i}^{0}-t_{ip}^{0}\geq\varphi+\frac{\delta}{v_{i}^{0}}$, by Lemma 4,
$v_{i_{p}}(t)$ is strictly increasing, therefore, $z_{i,i_{p}}(t_{i}^{0}%
)=\int_{t_{ip}^{0}}^{t_{i}^{0}}v_{i_{p}}(t)dt>\int_{t_{ip}^{0}}^{t_{i}^{0}%
}v_{i_{p}}^{0}dt\geq\varphi v_{i}^{0}+\delta$, which implies the safety
constraint (\ref{Safety}) is strict at $t_{i}^{0}$. Since we have $v_{i}%
^{0}=v_{i_{p}}^{0}$, by Lemma 3, $t_{i}^{m}-t_{ip}^{m}=t_{i}^{0}-t_{ip}%
^{0}\geq\varphi+\frac{\delta}{v_{i}^{0}}$. By Assumption 2, $z_{i,i_{p}}%
(t_{i}^{m})=(t_{i}^{m}-t_{i_{p}}^{m})v_{i_{p}}(t_{i_{p}}^{m})$. By Lemma 2,
$a_{i}=a_{i_{p}}$, and by Lemma 4, $v_{i}(t_{i}^{m})=v_{i_{p}}(t_{i_{p}}^{m}%
)$, therefore, $z_{i,i_{p}}(t_{i}^{m})=(t_{i}^{m}-t_{i_{p}}^{m})v_{i}%
(t_{i}^{m})\geq\varphi v_{i}(t_{i}^{m})+\frac{\delta}{v_{i}^{0}}v_{i}%
(t_{i}^{m})>\varphi v_{i}(t_{i}^{m})+\delta$. The safety constraint
(\ref{Safety}) is also strict at $t_{i}^{m}$. Because $a_{i}=a_{i_{p}}$ and
recalling that $u_{i}(t_{i}^{m})=-\lambda_{i}^{v}(t_{i}^{m})=0$, hence
$u_{i}(t_{i}^{m})=u_{i_{p}}(t_{i_{p}}^{m})=0$, CAVs $i$ and $i_{p}$ have the
same control law in the CZ, which implies they will take the same time to
arrive at the same point with the same speed in the CZ. Now, considering any
time instant $\tau_{i}\in(t_{i}^{0},t_{i}^{m})$ and $\tau_{i_{p}}\in(t_{i_{p}%
}^{0},t_{i_{p}}^{m})$ such that $\tau_{i}-t_{i}^{0}=\tau_{i_{p}}-t_{i_{p}}%
^{0}$, we have $\tau_{i}-\tau_{i_{p}}=t_{i}^{0}-t_{i_{p}}^{0}\geq\varphi
+\frac{\delta}{v_{i}^{0}}$. Because $v_{i_{p}}(t)$ is strictly increasing, it
follows that $z_{i,i_{p}}(\tau_{i})=\int_{\tau_{i_{p}}}^{\tau_{i}}v_{i_{p}%
}(t)dt>(\tau_{i}-\tau_{i_{p}})v_{i_{p}}(\tau_{i_{p}})$. Because $v_{i}%
(\tau_{i})=v_{i_{p}}(\tau_{i_{p}})$, then, $z_{i,i_{p}}(\tau_{i})>(\tau
_{i}-\tau_{i_{p}})v_{i}(\tau_{i})\geq\varphi v_{i}(\tau_{i})+\frac{\delta
}{v_{i}^{0}}v_{i}(\tau_{i})>\varphi v_{i}(\tau_{i})+\delta$ and the safety
constraint (\ref{Safety}) is always strict, i.e., $z_{i,i_{p}}(t)>\varphi
v_{i}(t)+\delta$ for all $t\in\lbrack t_{i}^{0},t_{i}^{m}]$.

Next, we consider the case $\beta>0$ and $v_{i}^{0}<v_{i_{p}}^{0}$. Suppose
there are two vehicles $i$ and $j$ such that $t_{i}^{0}=t_{j}^{0}$ and
$v_{i}^{0}<v_{j}^{0}$, and both use the optimal controller (\ref{Optimal_u}).
By Lemma 6, $a_{i}<a_{j}<0$, $v_{i}(t_{i}^{m})<v_{j}(t_{j}^{m})$ and
$t_{i}^{m}>t_{j}^{m}$. Because $u_{i}(t_{i}^{m})=0$ and $u_{i}(t)=a_{i}%
t+b_{i}$, we get $u_{i}(t)=a_{i}(t-t_{i}^{m})$. Similarly, $u_{j}%
(t)=a_{j}(t-t_{j}^{m})$. If $t=t_{i}^{m}$, because $t_{i}^{m}>t_{j}^{m}$, then
$u_{i}(t_{i}^{m})=0>a_{j}(t_{i}^{m}-t_{j}^{m})=u_{j}(t_{i}^{m})$. If
$t<t_{i}^{m}$, because $t_{i}^{m}>t_{j}^{m}$, then $t-t_{i}^{m}<t-t_{j}^{m}$
and $t-t_{i}^{m}<0$. Because $a_{i}<a_{j}<0$, then $a_{i}(t-t_{i}^{m}%
)>a_{j}(t-t_{j}^{m})$, thus $u_{i}(t)>u_{j}(t)$ for all $t\in\lbrack t_{i}%
^{0},t_{i}^{m}]$. Because $v_{i}^{0}<v_{j}^{0}$ and $v_{i}(t_{i}^{m}%
)<v_{j}(t_{j}^{m})$, then the speed curves of vehicles $i$ and $j$ will never
intersect, i.e., $v_{i}(t)<v_{j}(t)$ for all $t\in\lbrack t_{i}^{0},t_{i}%
^{m}]$, otherwise, there will be some time such that $u_{j}(t)\geq u_{i}(t)$,
which contradicts $u_{j}(t)<u_{i}(t)$ for all $t\in\lbrack t_{i}^{0},t_{i}%
^{m}]$. Now, considering the vehicle $j$ to be the case such that $v_{j}%
(t_{j}^{m})=v_{i_{p}}(t_{i_{p}}^{m})$ and $t_{j}^{0}-t_{i_{p}}^{0}\geq
\varphi+\frac{\delta}{v_{j}^{0}}$, then the safety constraint (\ref{Safety})
of $j$ will be satisfied for all $t\in\lbrack t_{j}^{0},t_{j}^{m}]\cup
(t_{j}^{m},t_{i}^{m}]$ following from the last paragraph and Assumption 2,
i.e., $z_{j,i_{p}}(t)>\varphi v_{j}(t)+\delta$. Because $v_{i}(t)<v_{j}(t)$
for all $t\in\lbrack t_{i}^{0},t_{i}^{m}]$, then $z_{i,i_{p}}(t)>z_{j,i_{p}%
}(t)$, hence $z_{i,i_{p}}(t)>z_{j,i_{p}}(t)>\varphi v_{j}(t)+\delta>\varphi
v_{i}(t)+\delta$. Therefore, $z_{i,i_{p}}(t)>\varphi v_{i}(t)+\delta$ for all
$t\in\lbrack t_{i}^{0},t_{i}^{m}]$.$\qquad\qquad\qquad\qquad\qquad\qquad
\qquad\qquad\qquad\qquad\;\;\blacksquare$

\textbf{Remark 1:} The significance of Theorem 1 is in ensuring that the
safety constraint (\ref{Safety}) is strict for all $t\in\lbrack t_{i}%
^{0},t_{i}^{m}]$ when $\beta>0$, $v_{i}^{0}\leq v_{i_{p}}^{0}$, $t_{i}%
^{0}-t_{ip}^{0}\geq\varphi+\frac{\delta}{v_{i}^{0}}$ and the optimal control
(\ref{Optimal_u}) is applied to $i$ and $i_{p}$. Therefore, in this case we do
not need to consider the safety constraint throughout the optimal trajectory, a fact
which significantly reduces computation. In contrast, when these conditions
are not satisfied, we need to consider the possibility of constrained arcs on
the optimal trajectory where $z_{i,i_{p}}(t)=\varphi v_{i}(t)+\delta$. This
case is discussed in the next subsection.

\begin{figure}[thpb]
	\centering
	\includegraphics[scale=0.5]{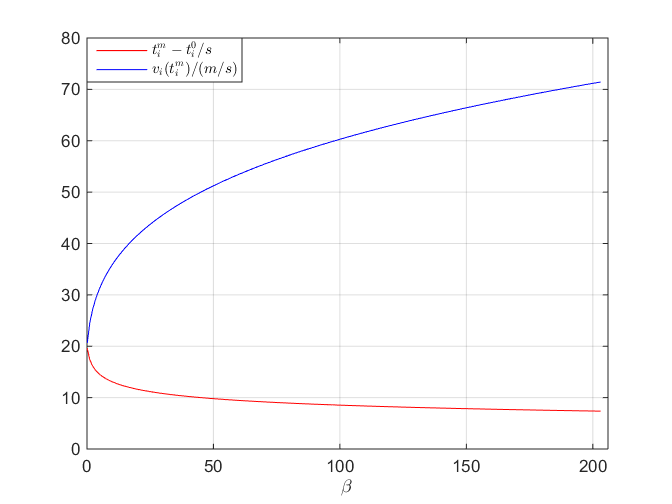}
	\caption{Optimal solutions for $\beta$ variation ($i-1 = i_p$). }
	\label{AbetaF}
\end{figure}

\begin{figure}[thpb]
	\centering
	\includegraphics[scale=0.5]{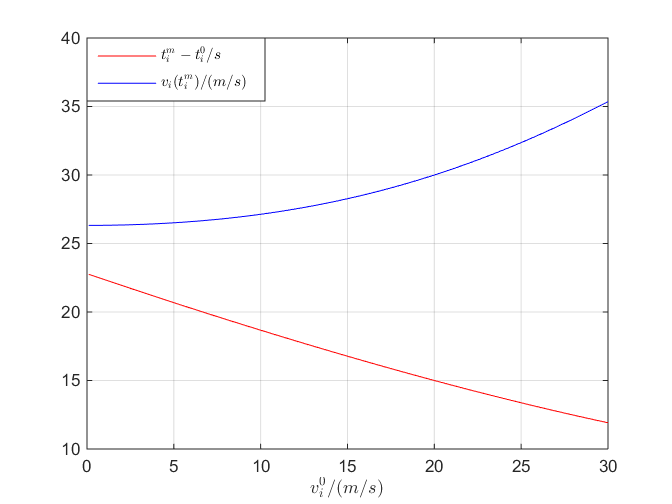}
	\caption{Optimal solutions for $v_i^0$ variation ($i-1 = i_p$).}
	\label{Av0F}
\end{figure}

\textbf{Numerical Example}: We have conducted simulations to solve (\ref{OptimalSolInA}) in MATLAB to evaluate the travel time and $v_i(t_i^m)$ when we change $\beta$ (or $\alpha$) and $v_i^0$. As $\beta$ varies with $v_i^0 = 20m/s, t_i^0 = 0s, L = 400$, the result is shown in Fig.\ref{AbetaF}. The result of changing the initial speed $v_i^0$ is shown in Fig.\ref{Av0F}, with $\beta = 2.667$ ($\alpha = 0.26$ when $u_{max} = -u_{min} = 0.4\times9.81$), $t_i^0 = 0s, L = 400$.

\subsubsection{\textbf{Safety Constraint Active}}
\label{sec:constrained_A}

When Theorem 1 does not apply, we must check whether the safety constraint
(\ref{Safety}) between vehicles $i$ and $i_{p}$ is ever violated for some
$t\in\lbrack t_{i}^{0},t_{i}^{m}]$ when they are under the optimal control
(\ref{Optimal_u}). If (\ref{Safety}) is violated, then we proceed as follows.

Suppose the safety constraint (\ref{Safety}) becomes active on an optimal trajectory at some time
$t_{1}\in(t_{i}^{0},t_{i}^{m}]$ (where $t_{1}$ will be optimally determined),
i.e., defining%
\begin{equation}
g_{i}(t):=x_{i}(t)+\varphi v_{i}(t)+\delta-x_{i_{p}}(t).\label{GFunction}%
\end{equation}
we have $g_{i}(t)<0$ for $t\in\lbrack t_{i}^{0},t_{1})$ and $g_{i}(t_{1})=0$.
Taking a time derivative, we get
\begin{equation}
\frac{dg_{i}(t)}{dt}=v_{i}(t)+\varphi u_{i}(t)-v_{i_{p}}%
(t)=0\label{GDerivative}%
\end{equation}
and it follows that over an optimal trajectory arc such that $g_{i}(t)=0$, the
optimal control is
\begin{equation}
u_{i}^{\ast}(t)=\frac{v_{i_{p}}(t)-v_{i}^{\ast}(t)}{\varphi},\text{ \ }t\geq
t_{1}\label{NewOptimal_u}%
\end{equation}
therefore,
\begin{equation}
\dot{v}_{i}^{\ast}(t)=\frac{v_{i_{p}}(t)-v_{i}^{\ast}(t)}{\varphi},\text{
	\ }t\geq t_{1}.\label{NewOptimal_v}%
\end{equation}

Clearly, if the original unconstrained optimal trajectory obtained through
(\ref{Optimal_u}), (\ref{Optimal_v}), (\ref{Optimal_x}) and (\ref{TransEqInA})
violates (\ref{Safety}) at any $t\in\lbrack t_{i}^{0},t_{i}^{m}]$ with
$t_{i}^{m}$ evaluated through (\ref{TransEqInA}), then \emph{a new optimal
	trajectory needs to be derived over the entire interval }$[t_{i}^{0},t_{i}%
^{m}]$. This is done by decomposing this trajectory into an initial segment
$[t_{i}^{0},t_{1})$ (where $t_{1}$ is to be determined as part of the
optimization process) followed by an arc where (\ref{Safety}) is active. 

Let us first assume that this arc applies over $[t_{1},t_{i}^{m}]$ and we proceed
as follows. We first solve the optimal control problem over $[t_{i}^{0}%
,t_{1})$ with initial conditions $x_{i}(t_{i}^{0}),v_{i}(t_{i}^{0})$ and the
terminal constraint $g_{i}(t_{1})=0$ together with the constraints
(\ref{Safety}), (\ref{VehicleConstraints}). In this solution, we treat $t_{1}$
as a parameter and obtain a solution dependent on $t_{1}$. We will then derive
the optimal value of $t_{1}$.

Let us assume (\ref{VehicleConstraints}) are inactive, as we did in obtaining
(\ref{Optimal_u}). Moreover, (\ref{Safety}) is inactive since we have assumed
it becomes active at some $t_{1}>t_{i}^{0}$. We can, therefore, derive
$x_{i}^{\ast}(t),v_{i}^{\ast}(t),u_{i}^{\ast}(t)$ (all functions of $t_{1}$)
which are similar to (\ref{Optimal_u})-(\ref{Optimal_x}) for all $t\in\lbrack
t_{i}^{0},t_{1})$. 

Let $\bm x_{i}:=(x_{i},v_{i})^T$,
$\bm\lambda_{i}:=(\lambda_{i}^{x},\lambda_{i}^{v})^T$. Following the notation and analysis of state inequalities in \cite{Bryson1969}, we write the state inequality constraint as $S_i(\bm x(t)):=x_{i}(t)+\varphi
v_{i}(t)+\delta-x_{i_{p}}(t)\leq 0$ and its first derivative as $S_i^{(1)}(\bm x(t), u_i(t)) = v_i(t) + \varphi u_i(t) - v_{i_p}(t)$. The new Hamiltonian is 
\begin{equation}H_{i}(\bm x_{i},\bm\lambda
_{i},u_{i})=\frac{1}{2}u_{i}^{2}+\lambda_{i}^{x}v_{i}+\lambda_{i}^{v}%
u_{i} + \beta + \mu S_i^{(1)},
\end{equation}
for $t\in[t_1,t_i^m]$. The tangency constraint is $N_{i}(\bm x_{i}(t_{1})):=x_{i}(t_{1})+\varphi
v_{i}(t_{1})+\delta-x_{i_{p}}(t_{1})=0$. 

Following \cite{Bryson1969}, at the entry point $t_1$, we have
\begin{equation} \label{LambdaCons}%
\left.\bm\lambda_i^T (t_1^-) = \bm\lambda_i^T (t_1^+) + \pi \frac{\partial N_i}{\partial \bm x_i}\right\vert_{t=t_1}, 
\end{equation}
\begin{equation} \label{HCons}%
\left.H_i(t_1^-) = H_i(t_1^+) - \pi \frac{\partial N_i}{\partial t}\right\vert
_{t=t_1}. 
\end{equation}
where $\pi$ is a constant Lagrange multiplier.

By (\ref{LambdaCons}), we have
\begin{equation} \label{LambdaConsE1}%
\lambda_i^x (t_1^-) = \lambda_i^x (t_1^+) + \pi, 
\end{equation}
\begin{equation} \label{LambdaConsE2}%
\lambda_i^v (t_1^-) = \lambda_i^v (t_1^+) + \pi\varphi, 
\end{equation}
and by (\ref{HCons}), we have
$$
\begin{aligned}
\frac{1}{2}u_i^2(t_1^-) + \lambda_i^x(t_1^-)v_i(t_1) + \lambda_i^v(t_1^-)u_i(t_1^-) = \frac{1}{2}u_i^2(t_1^+) \\+ \lambda_i^x(t_1^+)v_i(t_1) + \lambda_i^v(t_1^+)u_i(t_1^+)+ \pi v_{i_p}(t_1) .\end{aligned}
$$

Combining (\ref{LambdaConsE1}), the optimality condition $u_i(t_1^-) = -\lambda_i^v(t_1^-)$ and the last equation, we have
$$
\begin{aligned}
-\frac{1}{2}u_i^2(t_1^-) + \pi v_i(t_1)  = \\\frac{1}{2}u_i^2(t_1^+) + \lambda_i^v(t_1^+)u_i(t_1^+)+ \pi v_{i_p}(t_1), \end{aligned}
$$

 On the constrained arc, we have from (\ref{NewOptimal_u}): $u_i(t_1^+) = \frac{v_{i_p}(t_1) - v_i(t_1)}{\varphi}.$ Therefore, the last equation can be rewritten as
$$
-\frac{1}{2}u_i^2(t_1^-)  = \frac{1}{2}u_i^2(t_1^+) + \lambda_i^v(t_1^+)u_i(t_1^+)+ \varphi\pi u_{i}(t_1^+) 
$$

Combining (\ref{LambdaConsE2}) and the last equation, we have
$$
-\frac{1}{2}u_i^2(t_1^-)  = \frac{1}{2}u_i^2(t_1^+) + \lambda_i^v (t_1^-)u_{i}(t_1^+)
$$

Further by optimality condition $u_i(t_1^-) = -\lambda_i^v(t_1^-)$, the last equation can be rewitten as
$$
-\frac{1}{2}u_i^2(t_1^-)  = \frac{1}{2}u_i^2(t_1^+) - u_i(t_1^-)u_{i}(t_1^+)
$$

By simplifying the last equation, we get
\begin{equation}
u_i(t_1^-)  = u_i(t_1^+) \label{InteriorPointConstraint}%
\end{equation}

Recall that the optimal solution for $t\in\lbrack t_{i}^{0},t_{1})$ is given by
\begin{equation}
\begin{aligned} &u_i(t) = a_it+b_i\\ &v_i(t) = \frac{1}{2}a_it^2 + b_it + c_i\\ &x_i(t) = \frac{1}{6}a_it^3 + \frac{1}{2}b_it^2 + c_it + d_i \end{aligned} \label{OptimalSolBeforeT1}%
\end{equation}

On the constrained arc, we can then solve (\ref{NewOptimal_v}) for the optimal solution $v_{i}^{\ast
}(t)$ with initial condition $v_{i}(t_{1})$ known from
(\ref{OptimalSolBeforeT1}), and hence obtain $x_{i}^{\ast}(t)$ with initial
condition $x_{i}(t_{1})$ from (\ref{OptimalSolBeforeT1}). Suppose $i_p$ is under unconstrained optimal control (\ref{Optimal_u}), $v_{i_p}^*(t)$ is known to CAV $i$. Moreover, by Assumption 2, we know that $v_{i_p}^*(t)$ is a constant over $[t_{i_p}^m, t_i^m]$. Therefore, we need to divide the solution over two intervals, i.e., the explicit solution for CAV $i$ is:
\begin{equation}
\begin{aligned}
&x_i^*(t) = 
\left\{\!
\begin{array}{lcl}
\begin{aligned}&-c_{v_1}\varphi e^{-\frac{1}{\varphi}t} + x_{i_p}^*(t) - d_{i_p}-\\&\varphi (v_{i_p}^*(t) - c_{i_p}) + a_{i_p}\varphi^2t + c_{x_1}\end{aligned} &t\in[t_1,t_{i_p}^m]\\
-c_{v_2}\varphi e^{-\frac{1}{\varphi}t} + v_{i_p}^*(t_{i_p}^m)t + c_{x_2} &t\in(t_{i_p}^m,t_i^m]
\end{array}
\right.\\
&v_i^*(t) = 
\left\{\!
\begin{array}{lcl}
c_{v_1}e^{-\frac{1}{\varphi}t}\! +\! v_{i_p}^*(t) \!-\!\varphi u_{i_p}^*(t)\! +\! a_{i_p}\varphi^2 &t\in[t_1,t_{i_p}^m]\\
c_{v_2}e^{-\frac{1}{\varphi}t} + v_{i_p}^*(t_{i_p}^m) &t\in(t_{i_p}^m,t_i^m]
\end{array}
\right.\\
&u_i^*(t) = 
\left\{
\begin{array}{lcl}
-\frac{c_{v_1}}{\varphi}e^{-\frac{1}{\varphi}t} + u_{i_p}^*(t) -\varphi a_{i_p} &t\in[t_1,t_{i_p}^m]\\
-\frac{c_{v_2}}{\varphi}e^{-\frac{1}{\varphi}t} &t\in(t_{i_p}^m,t_i^m]
\end{array}
\right.
\end{aligned}\label{NewOptimalUVX}%
\end{equation}
where $c_{x_1} = x_i^*(t_1) +c_{v_1}\varphi e^{-\frac{1}{\varphi}t_1} - x_{i_p}^*(t_1) + d_{i_p} +\varphi (v_{i_p}^*(t_1) - c_{i_p}) - a_{i_p}\varphi^2t_1$, $c_{x_2} = x_i^*(t_{i_p}^m) + c_{v_2}\varphi e^{-\frac{1}{\varphi}t_{i_p}^m} - v_{i_p}^*(t_{i_p}^m)t_{i_p}^m$, $c_{v_1} = e^{\frac{1}{\varphi}t_1}(v_{i}^*(t_1) - v_{i_p}^*(t_1) +\varphi u_{i_p}^*(t_1)- a_{i_p}\varphi^2)$, $c_{v_2} = e^{\frac{1}{\varphi}t_{i_p}^m}(v_i^*(t_{i_p}^m) - v_{i_p}^*(t_{i_p}^m))$.
If $i_p$ is also under constrained optimal control, the optimal solution for $i$ is recursively determined by (\ref{NewOptimal_v}) starting from the first vehicle that is under unconstrained optimal control.

The value of the entry point $t_1$ can be directly obtained by combining (\ref{InteriorPointConstraint}), initial conditions, terminal conditions and the tangency constraint $N_i(\bm x_i(t_1))$, i.e., we have the following algebraic equations
\begin{equation}
\begin{aligned}
&a_it_1 + b_i  = \frac{v_{i_p}(t_1) - v_i(t_1)}{\varphi},\\
&\frac{1}{2}a_i\cdot(t_i^0)^2 + b_it_i^0 + c_i = v_i^0,\\ &\frac{1}{6}a_i\cdot(t_i^0)^3 + \frac{1}{2}b_i\cdot(t_i^0)^2 + c_it_i^0+d_i = 0,\\
&x_i(t_1) +\varphi v_i(t_1) + \delta = x_{i_p}(t_1),\\
&\varphi v_i(t_i^m) + \delta = v_{i_p}(t_{i_p}^m)(t_i^m - t_{i_p}^m),\\
&x_{i}(t_i^m) = L.
\end{aligned}\label{InteriorSol}%
\end{equation}
to solve for $a_i,b_i,c_i,d_i,t_1,t_i^m$.

\begin{figure}[thpb]
	\centering
	\includegraphics[scale=0.5]{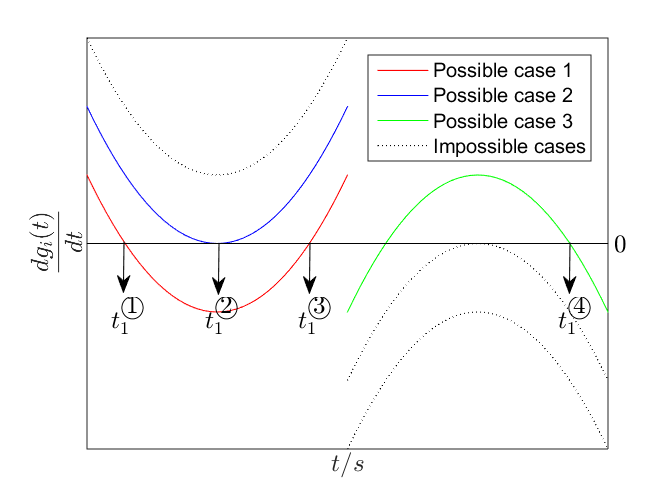}
	\caption{The three possible cases for $\frac{dg_i(t)}{dt}$. If $t_1$ locates at $t_1^{\textcircled{3}}$, then $t_1$ is not the first time such that the safety constraint (\ref{Safety}) becomes active. Otherwise, $t_1$ is indeed the first time.}
	\label{fig:cases}%
\end{figure}

In what follows, we first assume that $i_p$ is under unconstrained optimal control. Solving (\ref{InteriorSol}) generally provides multiple solutions for $t_1$, some of which may not be feasible. Since we have assumed $i_p$ is under unconstrained optimal control, we know that $\frac{dg_i(t)}{dt}$ is a quadratic funtion, and there are total six cases as shown in Fig. \ref{fig:cases}. By (\ref{InteriorPointConstraint}), we have $\frac{dg_i(t)}{dt}\vert_{t = t_1} = 0$, thus, $\frac{dg_i(t)}{dt}$ must intersect with time axis at $t_1$. We can, therefore, exclude these two cases where $\frac{dg_i(t)}{dt}$ does not intersect the time axis as shown in Fig. \ref{fig:cases}. By Assumption 1, the safety constraint is strict at $t_i^0$, therefore, we have $g_i(t_i^0) < 0$, $g_i(t)$ cannot decrease for all $t\in[t_i^0,t_1)$ such that $g_i(t_1) = 0$, so we can exclude another case that is also shown in Fig. \ref{fig:cases}. Now, we have three cases for $\frac{dg_i(t)}{dt}$, if $t_1$ locates at $t_1^{\textcircled{3}}$ shown in Fig.\ref{fig:cases}, then $t_1$ is not the first time such that the safety constraint (\ref{Safety}) becomes active, which is infeasible. The remaining possible locations for $t_1$ shown in Fig. \ref{fig:cases} are feasible.

However, as we can see in (\ref{NewOptimalUVX}), the last two equations in (\ref{InteriorSol}) are exponential functions of time as we have already assumed the constrained arc (\ref{NewOptimalUVX}) has no exit point. Consequently, they are hard to solve directly. This motivates an alternative approach in which we first solve the first four equations in (\ref{InteriorSol}) for $a_i, b_i, c_i,d_i$ in terms of $t_1$:
\begin{equation}\label{OptimalSol}%
\begin{aligned} &a_it_1 + b_i  = \frac{v_{i_p}(t_1) - v_i(t_1)}{\varphi},\\
&\frac{1}{2}a_i\cdot(t_i^0)^2 + b_it_i^0 + c_i = v_i^0,\\ &\frac{1}{6}a_i\cdot(t_i^0)^3 + \frac{1}{2}b_i\cdot(t_i^0)^2 + c_it_i^0+d_i = 0,\\
&x_i(t_1) +\varphi v_i(t_1) + \delta = x_{i_p}(t_1). \end{aligned}
\end{equation}

Similarly as in (\ref{InteriorSol}), the optimal control (\ref{OptimalSolBeforeT1}) solved by (\ref{OptimalSol}) for $t\in\lbrack t_{i}^{0},t_{1})$ cannot guarantee that $t_1$ is the first time such that the safety constraint (\ref{Safety}) becomes active (if that happens, then $t_1$ is infeasible). Therefore, we need to exclude such infeasible $t_1$ as explained next.

Under Assumption 1, there may exist some cases such that $v_i(t_1^-) < v_{i_p}(t_1^-)$ when the safety constraint (\ref{Safety}) becomes active between CAV $i$ and $i_p$. By Assumption 1, it follows that $g_i(t_i^0) < 0$. We also have $g_i(t_1) = 0$. However, the sign of the derivative $\left.\frac{dg_i(t)}{dt}\right\vert_{t = t_1^-}$ is unknown. If $\left.\frac{dg_i(t)}{dt}\right\vert_{t = t_1^-} \geq 0$ i.e., $v_{i}(t_1^-)+\varphi u_{i}(t_1^-)\geq v_{i_{p}}(t_1^-)$, it is possible that $v_i(t_1^-) < v_{i_p}(t_1^-)$ when $ u_{i}(t_1) > 0$. Similarly, if $\left.\frac{dg_i(t)}{dt}\right\vert_{t = t_1^-} < 0$, there is also a possibility that $v_i(t_1^-) < v_{i_p}(t_1^-)$. This property is helpful to understand the process of finding the infeasible set for $t_1$.

\vspace{2ex} \textbf{Definition 1: } We define a set $I_i$ as:
\begin{equation}\label{eqn:infeasiblet1}%
\begin{aligned}
I_i:=&\{t_1\in (t_i^0,t_i^m] |u_i(t_1) + \varphi a_i > u_{i_p}^*(t_1)\},
\end{aligned}
\end{equation}
where $u_i(t_1)$ is from (\ref{OptimalSolBeforeT1}).

\vspace{2ex} \textbf{Lemma 7: }Under Assumption 1, if $I_i$ is non-empty, then any $ t_1\in I_i$ is not the first time such that the safety constraint (\ref{Safety}) becomes active. \vspace{2ex}

$\emph{Proof:}$ Since the safety constraint (\ref{Safety}) becomes active at $t_1\in I$, it follows that $g_i(t_1) = 0$. By the first equation of (\ref{InteriorSol}), we have $\left.\frac{dg_i(t)}{dt}\right\vert_{t = t_1^-} = 0$. If $u_i(t_1) + \varphi a_i > u_{i_p}(t_1)$, i.e., $\left.\frac{d^2g_i(t)}{dt^2}\right\vert_{t = t_1^-} > 0$, then the function $g_i(t)\rightarrow 0$ as $t\rightarrow t_1$ from the positive side. By the continuity of $g_i(t)$, we know that the safety constraint is violated for some $t<t_1$. By Assumption 1, the safety constraint (\ref{Safety}) is initially strict, thus, there exist time instant $t_a < t_1$ such that $g_i(t_a) = 0$. Therefore, any $ t_1\in \{t_1\in (t_i^0,t_i^m] |u_i(t_1) + \varphi a_i > u_{i_p}^*(t_1)\}$ is not the first time such that the safety constraint (\ref{Safety}) becomes active.$\;\;\;\;\blacksquare$

We know that $t_1\in I_i$ is infeasible since these $t_1$ will make the safety constraint (\ref{Safety}) become violated for a time interval. Therefore, we need to exclude $I_i$ for the safety constraint active case.

\vspace{2ex} \textbf{Theorem 2: }Under Assumption 1, $g_i(t_1) = 0$ and $\left.\frac{dg_i(t)}{dt}\right\vert_{t = t_1^-}\geq 0$ are the necessary conditions for $t_1$ to be the first time such that the safety constraint (\ref{Safety}) becomes active. Moreover, if CAV $i_p$ is under unconstrained optimal control (\ref{Optimal_u}), $g_i(t_1) = 0$, $\left.\frac{dg_i(t)}{dt}\right\vert_{t = t_1^-} = 0$ and $\left.\frac{d^2g_i(t)}{dt^2}\right\vert_{t = t_1^-} \leq 0$ are sufficient conditions. \vspace{2ex}

$\emph{Proof:}$  If $t_1$ is the first time such that the safety constraint (\ref{Safety}) becomes active, then it follows that $g_i(t_1) = 0$. By Assumption 1, we have $g_i(t_i^0) < 0$. If $\left.\frac{dg_i(t)}{dt}\right\vert_{t = t_1^-} < 0$, then we have $g_i(t_1^-) > 0$. By the continuity of $g_i(t)$, it follows that we have another time instant $t_a < t_1$ such that the safety constraint (\ref{Safety}) becomes active, which contradicts the fact that $t_1$ is the first time. Therefore, we have $\left.\frac{dg_i(t)}{dt}\right\vert_{t = t_1^-} \geq 0$, thus, $g_i(t_1) = 0$ and $\left.\frac{dg_i(t)}{dt}\right\vert_{t = t_1^-}\geq 0$ are the necessary conditions. 

By (\ref{OptimalSolBeforeT1}), it follows that $v_i(t)$ is a second order polynomial funtion of time. Since CAV $i_p$ is under unconstrained optimal control (\ref{Optimal_u}), it follows from (\ref{Optimal_v}) that $v_{i_p}^*(t)$ is also a second order polynomial function of time for $t\in[t_{i_p}^0, t_{i_p}^m]$ and $v_{i_p}^*(t)= v_{i_p}^*(t_{i_p}^m)$ for $t\in[t_{i_p}^m, t_{i}^m]$ following from Assumption 1. Therefore, $\frac{dg_i(t)}{dt}$ is a second order polynomial funtion of time for $t\in[t_{i}^0, t_1)$. By Lemma 7, $\left.\frac{d^2g_i(t)}{dt^2}\right\vert_{t = t_1^-} \leq 0$ indicates that $t_1\notin I_i$, and further by $\left.\frac{dg_i(t)}{dt}\right\vert_{t = t_1^-} = 0$, we have $\frac{dg_i(t)}{dt} > 0, \forall t\in[t_i^0,t_1)$ or $\frac{dg_i(t)}{dt}$ is negative for $t\in[t_i^0, t_a)$ and becomes positive for $t\in(t_a, t_1)$ (where $t_a\in(t_i^0,t_1)$). By Assumption 1, the safety constraint (\ref{Safety}) is strict at $t_i^0$. Thus, $g_i(t) < 0, \forall t\in[t_i^0,t_1)$ and $t_1$ is the only time such that the safety constraint (\ref{Safety}) becomes active for $t\in[t_i^0,t_1]$. Therefore, if CAV $i_p$ is under unconstrained optimal control (\ref{Optimal_u}), $g_i(t_1) = 0$, $\left.\frac{dg_i(t)}{dt}\right\vert_{t = t_1^-} = 0$ and $\left.\frac{d^2g_i(t)}{dt^2}\right\vert_{t = t_1^-} < 0$ are sufficient conditions for $t_1$ to be the first time that the safety constraint (\ref{Safety}) becomes active. $\;\;\;\blacksquare$

\textbf{Remark 2:} Theorem 2 applies only to the case where $i_p$ is not under constrained optimal control, i.e., the safety constraint (\ref{Safety}) never became active. If this does not hold, then the form of $\frac{dg_i(t)}{dt}$ is no longer quadratic, in which case we need to identify the set $I_i$ by determining all $t_1\in (t_i^0,t_i^m]$ such that there exists $ t\in[t_i^0,t_1)$ that $g_i(t) > 0$. Clearly, the computation effort for fully determining $I_i$ is more intensive in such cases.

Theorem 2 provides simple to check conditions to find all feasible $t_1$ that are the first time such that the safety constraint (\ref{Safety}) becomes active. Otherwise, we need to do more computation to decide whether $t_1$ is feasible or not, as suggested in Remark 2.

Recall that we have assumed that there is no exit point from this constraint arc prior to $t_{i}^{m}$. Using the optimal solutions (\ref{OptimalSolBeforeT1}) for $[t_{i}%
^{0},t_{1})$ and (\ref{NewOptimal_u}) for $[t_{1},t_{i}^{m}]$ in
(\ref{CAV_problem}), we obtain the optimal value of the objective function
$J_{i}^{\ast}(t_{1})$ parameterized by $t_{1}$: {\small
	\begin{equation}
	\begin{aligned} J_i^*(t_1)\! =\! \beta(t_i^m\! -\! t_i^0) \!+\! \frac{a_i^2}{6}(t_1^3 \!-\! (t_i^0)^3) \!+\! \frac{1}{2}a_ib_i(t_1^2 \!-\! (t_i^0)^2) \\+ \frac{1}{2}b_i^2(t_1 - t_i^0) + \int_{t_1}^{t_{i_p}^m}\frac{1}{2}(u_i^*(t))^2 dt + \int_{t_{i_p}^m}^{t_i^m}\frac{1}{2}(u_i^*(t))^2 dt \end{aligned}\label{J1*}%
	\end{equation}
} where $u_{i}^{\ast}(t)$ is the optimal control from (\ref{NewOptimal_u}) and
depends on $t_{1}$ as do the constants $a_{i},b_{i}$ above. $u_{i}^{\ast}(t)$ is also a function of $t_{1}$ for $t\in\lbrack t_{1},t_{i_{p}}^{m}]$ as its explicit solution shown in (\ref{NewOptimalUVX}). The
optimal solution for $t_{1}$ is obtained by finding $t_{1}$ that
minimizes $J_{i}^{\ast}(t_{1})$. Note that the value of
$t_{i}^{m}$ is obtained by setting $x_{i}^{\ast}(t_{i}^{m})=L$. If we apply the optimal controller
(\ref{NewOptimal_u}) till $t_{i}^{m}$, then $t_{i}^{m}$ is also
dependent on $t_{1}$ as the explicit solution of $x_{i}^{\ast}(t_{i}^{m})$ shown in (\ref{NewOptimalUVX}).

If $t_1^* = t_i^m$, then the interior point $t_1$ degenerates to a terminal point. We then need to take the safety constraint as a terminal boundary constraint and solve a new optimal control problem that will be discussed in Sec.\ref{sec:Nequal}.

Let us now explore the case where there exists an exit point from the
constraint arc (\ref{NewOptimal_u}) prior to $t_{i}^{m}$. First, observe that
$t_{1}$ is the first instant when CAV $i$ catches up with $i_{p}$ so as to
activate (\ref{Safety}), therefore $v_{i}(t_{1})\geq v_{i_{p}}(t_{1})$. It is
easy to see that if $u_{i}^{\ast}(t)$ in (\ref{NewOptimal_u}) remains
negative, then (\ref{NewOptimal_u}) remains the optimal solution. However, if
$v_{i_{p}}(t_{2})>v_{i}(t_{2})$ at some $t_{2}\in\lbrack t_{1},t_{i}^{m}]$,
this means that it is possible (\ref{NewOptimal_u}) is no longer optimal
because the safety constraint (\ref{Safety}) may become inactive again. In
this case, we need to solve another optimal control problem similar to that of
the no-active-constraint case (\ref{Optimal_u})-(\ref{Optimal_x}) but with
initial condition $x_{i}(t_{2})$ obtained from the solution of
(\ref{NewOptimal_v}), with the same terminal conditions as in
(\ref{CAV_problem}), subject to (\ref{VehicleDynamics}), (\ref{Safety}) and
(\ref{VehicleConstraints}), and with $t_{i}^{m}$ once again a free terminal
time. For the new arc starting at $t_{2}$, we can solve for $a_{i},b_{i}%
,c_{i},d_{i},t_{i}^{m}$ and $t_{2}$ similar to (\ref{OptimalSolInA}) using
\begin{equation}
\begin{aligned} &\frac{1}{2}a_it_2^2 + b_it_2 + c_i = v_i^*(t_2),\\ &\frac{1}{6}a_it_2^3 + \frac{1}{2}b_it_2^2 + c_it_2+d_i = x_i^*(t_2),\\ &\frac{1}{6}a_i\cdot(t_i^m)^3 + \frac{1}{2}b_i\cdot(t_i^m)^2 + c_it_i^m+d_i = L,\\ &a_it_i^m + b_i = 0,\\ &\beta + \frac{1}{2}a_i^2\cdot(t_i^m)^2 + a_ib_it_i^m + a_ic_i = 0,\\ &a_it_2 + b_i = u_i^*(t_2) \end{aligned}\label{SolveforT2inA}%
\end{equation}
where $x_{i}^{\ast}(t_{2})$, $v_{i}^{\ast}(t_{2})$, $u_{i}^{\ast}(t_{2})$ are
the optimal solutions from (\ref{NewOptimal_u})-(\ref{NewOptimal_v}) and the
last equation ensures the continuity of $u_{i}(t)$, otherwise, the safety
constraint (\ref{Safety}) is immediately violated. If a feasible solution for
$t_{2}$ exists in solving (\ref{SolveforT2inA}), then we evaluate the
objective (\ref{CAV_problem}) again as in (\ref{J1*}) in order to determine
the optimal values of $t_{1}$ and $t_{2}$; otherwise, the trajectory
determined above over $[t_{1},t_{i}^{m}]$ is optimal. 

If a feasible solution for $t_{2}$ is determined, it is possible that the safety constraint
(\ref{Safety}) becomes active again at some $t_{3}\in\lbrack t_{2},t_{i}^{m}%
]$. Thus, we use the same method to deal with the construction of a complete
optimal trajectory recursively.

In a nutshell, we can summarize the method of finding the optimal $t_1$ (or $t_2$ if it exists) and $x_{i}^*(t), v_{i}^*(t),  u_{i}^*(t)$ by Algorithm \ref{algo:1}, which includes all cases, including the case when $i_p$ is under constrained optimal control, or even recursively constrained optimal control.

\begin{algorithm}[t] 
	\caption{Safety constrained optimal trajectory, $i_p = i-1$} 
	\hspace*{0.02in} {\bf Input:} 
	 $t_i^0, v_i^0, x_{i_p}^*(t), v_{i_p}^*(t),  u_{i_p}^*(t)$\\
	\hspace*{0.02in} {\bf Output:} 
	$t_1^*, x_{i}^*(t), v_{i}^*(t),  u_{i}^*(t)$, $t_2^*$ (if it exists)
	\begin{algorithmic}[1] 
		\State solve (\ref{OptimalSol})  \label{algo:1}
		\If{CAV $i_p$ is under unconstrained optimal control (\ref{Optimal_u})} 		
		\State $I_i:=\{t_1\in (t_i^0,t_i^m] |u_i(t_1) + \varphi a_i > u_{i_p}^*(t_1)\}$
		\Else
		\State $I_i = \{t_1\in (t_i^0,t_i^m]|\exists t\in[t_i^0,t_1), g_i(t) > 0\}$
		\EndIf
		\State get feasible set $F_i:=(t_i^0,t_i^m)\setminus I_i$ for $t_1$
		\State solve (\ref{NewOptimalUVX}) or its recursive form if $i_p$ is under constrained optimal control		
		\State solve (\ref{SolveforT2inA})
		\If {(\ref{SolveforT2inA}) has no feasible solutions ($t_2$ does not exist)	}
		\State solve $t_i^m$ in terms of $t_1$ by $x_i^*(t_i^m) = L$.
		\State get $J_i^*(t_1)$ by (\ref{J1*})
		\State solve for $t_1^*$ over $F_i$
		\State result = $t_1^*, x_{i}^*(t), v_{i}^*(t),  u_{i}^*(t)$
		\Else
		\State get $J_i^*(t_1)$ by a similar form as (\ref{J1*})
		\State solve for $t_1^*$ over $F_i$ (and $t_2^*$)
		\State result = $t_1^*, x_{i}^*(t), v_{i}^*(t),  u_{i}^*(t) , t_2^*$
		\EndIf
		\If{$t_1^* = t_i^m$}
		\State take the safety constraint (\ref{Safety}) as a terminal boundary constraint as in Sec.\ref{sec:Nequal}
		\If {the safety constraint (\ref{Safety}) is satisfied $\forall t\in[t_i^0,t_i^m]$}
		\State  result = $t_1^*(= t_i^m), x_{i}^*(t), v_{i}^*(t),  u_{i}^*(t)$
		\EndIf
		\EndIf
		\State \Return result
	\end{algorithmic}
\end{algorithm}

The next theorem ensures that if an optimal trajectory includes an arc over
which the safety constraint (\ref{Safety}) is initially satisfied, then the optimal control (\ref{NewOptimal_u}) never violates the constraint (\ref{VehicleConstraints}). 

\textbf{Theorem 3: }If $u_{min}\leq\frac{v_{i_{p}}(t_{1})-v_{i}(t_{1}%
	)}{\varphi}\leq u_{max}$, then under optimal control (\ref{NewOptimal_u}) for $t\in\lbrack
t_{1},t_{i}^{m}]$\vspace{2ex}, $u_{min}\leq u_{i}^{\ast}(t)\leq u_{max}$.

$\emph{Proof:}$ Taking a time derivative in (\ref{NewOptimal_u}) we get
\begin{equation}
\dot{u}_{i}(t)=\frac{u_{i_{p}}(t)-u_{i}(t)}{\varphi},t\geq t_{1}.
\end{equation}
with $u_{i}(t_{1})\geq u_{min}$. There are three cases to consider.

\emph{Case 1}: $u_{i}(t_{1})=u_{min}$, so that $\dot{u}_{i}(t)=\frac{u_{i_{p}%
	}(t)-u_{min}}{\varphi}$. Because $u_{min}\leq u_{i_{p}}(t)\leq u_{max}$ on an
optimal trajectory for vehicle $i_{p}$, we get $\dot{u}_{i}(t)\geq0$, which
means $u_{i}(t)$ is non-decreasing. 

\emph{Case 2}: $u_{i}(t_{1})=u_{max}$, so that $\dot{u}_{i}(t)=\frac{u_{i_{p}%
	}(t)-u_{max}}{\varphi}$. Because $u_{min}\leq u_{i_{p}}(t)\leq u_{max}$ on an
optimal trajectory for vehicle $i_{p}$, we get $\dot{u}_{i}(t)\leq0$, which
means $u_{i}(t)$ is non-increasing.

\emph{Case 3}: $u_{min}<u_{i}(t_{1})<u_{max}$. In this case, we have $\dot
{u}_{i}(t)\geq\frac{u_{min}-u_{i}(t_{1})}{\varphi}$ and $\dot{u}_{i}(t)$ may
be negative, therefore, $u_{i}(t)$ is allowed to decrease when $u_{i}%
(t)>u_{min}$. But when $u_{i}(t)$ approaches $u_{min}$, the lower bound of
$\dot{u}_{i}(t)$ will approach zero and $u_{i}(t)$ is once again
non-decreasing, therefore, $u_{i}(t)\geq u_{min}$ for all $t\in\lbrack
t_{1},t_{i}^{m}]$. On the other hand, we also have $\dot{u}_{i}(t)\leq
\frac{u_{max}-u_{i}(t_{1})}{\varphi}$ and $\dot{u}_{i}(t)$ may be positive,
therefore, $u_{i}(t)$ is allowed to increase when $u_{i}(t)<u_{max}$. But when
$u_{i}(t)$ approaches $u_{max}$, the upper bound of $\dot{u}_{i}(t)$ will
approach zero, then $u_{i}(t)$ is once again non-increasing, therefore,
$u_{i}(t)\leq u_{max}$, $\forall t\in\lbrack t_{1},t_{i}^{m}]$. $\quad\blacksquare$

\begin{figure}[thpb]
	\centering
	\includegraphics[scale=0.5]{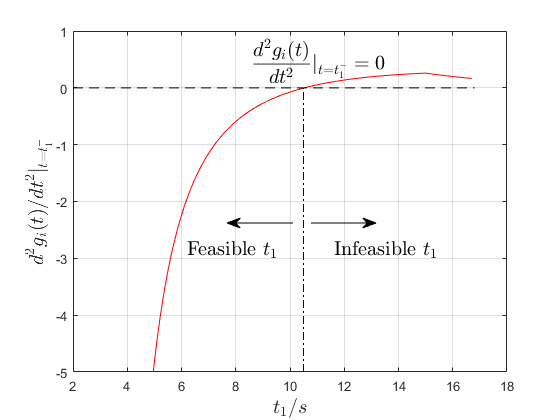}
	\caption{The second order derivative of $g_i(t)$ with respect to $t$ at $t = t_1^-$. }
    \label{fig:ddG}%
\end{figure}

\textbf{Numerical Example:} The initial parameters for $i$ and $i_p$ are with $i_{p}=i-1$, $t_{i_{p}}^{0}=0s$, $v_{i_{p}}%
^{0}=20m/s,$  $t_{i}^{0}=2.7s,$ $v_{i}^{0}=27m/s$, $\beta=2.667$
($\alpha=0.2573$), $\varphi = 1.8s$, $\delta = 0m$, $L = 400m$. If we apply (\ref{Optimal_u}), we know that the safety constraint (\ref{Safety}) will be violated. Therefore, we need to solve for the constrained optimal control. We use (\ref{OptimalSolBeforeT1}) for $t\in[t_i^0,t_1)$, (\ref{NewOptimalUVX}) for $t\in[t_1,t_2)$ and the optimal control solved by (\ref{SolveforT2inA}) for $t\in[t_2,t_i^m]$. Firstly, we check whether there is infeasible interval $I_i$ for $t_1$, as shown in Fig.\ref{fig:ddG}.

\begin{figure}[thpb]
	\centering
	\includegraphics[scale=0.5]{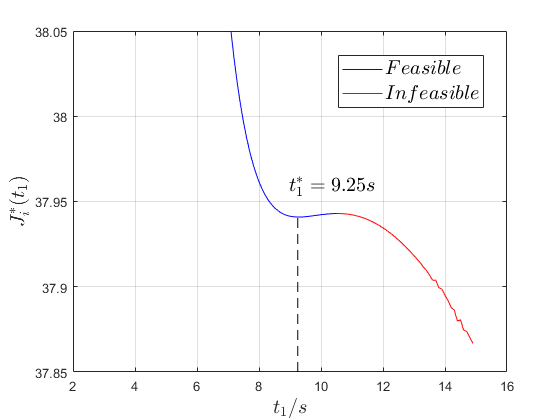}
	\caption{ $J_i^*(t_1)$ with respect to $t_1$. }
	\label{fig:J_A}%
\end{figure}

It follows from Fig.\ref{fig:ddG} that the infeasible interval $I_i = (10.5, t_i^m]$ does exist in this case, then $F_i = [t_i^0,t_i^m]\setminus I_i$. Because $v_i(t_1) < v_{i_p}(t_1)$, $\forall t_1\in F_i$, then $t_2$ exists following from Theorem 4. It follows from (\ref{SolveforT2inA}) that $t_2$ depends on $t_1$ and $t_i^m$ is free. The optimal objective function with respect to $t_1$ is shown in Fig.\ref{fig:J_A}, and we get $t_1^* = 9.25s$.

\begin{figure}[thpb]
	\centering
	\includegraphics[scale=0.5]{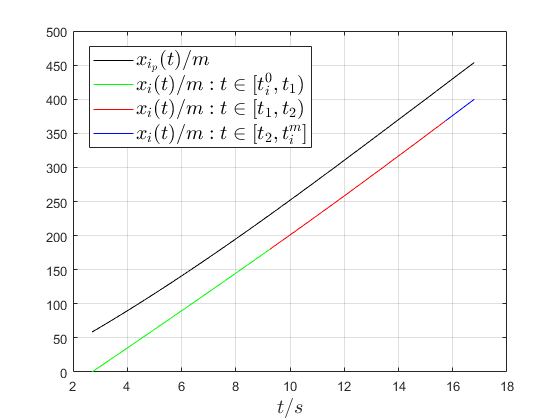}
	\caption{The position profile for $i$ and $i_p$. }
	\label{fig:pos_A}%
\end{figure}

\begin{figure}[thpb]
	\centering
	\includegraphics[scale=0.5]{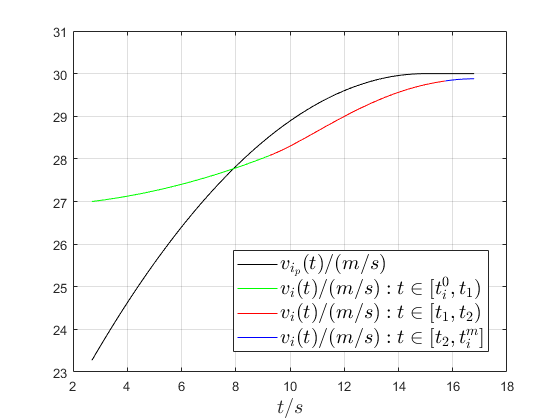}
	\caption{The speed profile for $i$ and $i_p$. }
	\label{fig:speed_A}%
\end{figure}

\begin{figure}[thpb]
	\centering
	\includegraphics[scale=0.5]{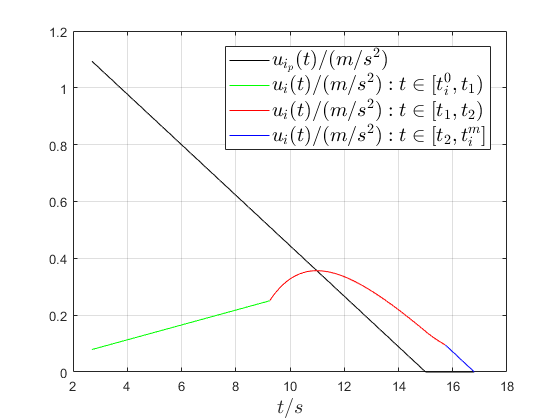}
	\caption{The control profile for $i$ and $i_p$. }
	\label{fig:control_A}%
\end{figure}

\begin{figure}[thpb]
	\centering
	\includegraphics[scale=0.5]{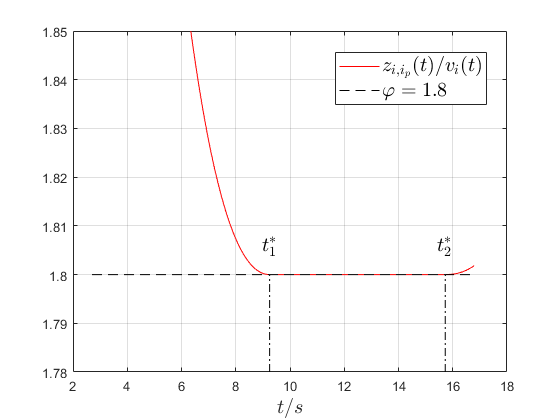}
	\caption{The safety profile for $i$ and $i_p$. }
	\label{fig:safety_A}%
\end{figure}

 We continue to study the state and safety constraint profiles at $t_1^* = 9.25s$ and $t_2^* = 15.76s$, as shown in Fig.\ref{fig:pos_A}-\ref{fig:safety_A}. 


\subsection{Decentralized Optimal Control when $i-1>i_{p}$}
\label{sec:Nequal}

In this case, CAV $i_{p}$ which physically precedes $i\in S(t)$ is
different from $i-1$ which, therefore, is in a different lane than $i$. This
implies that we need to consider the safe merging constraint
(\ref{SafeMerging}) at $t=t_{i}^{m}$. We define a new state vector $\bm
x_{i}(t):=(x_{i}(t),v_{i}(t))^T$. We also
define a new terminal constraint $\psi_{i,2}(\bm x_{i}(t_{i}^{m}),t_{i}%
^{m}):=x_{i}(t_{i}^{m})+\varphi v_{i}(t_{i}^{m})+\delta-x_{i-1}(t_{i}^{m})=0$,
where we have replaced the inequality in (\ref{SafeMerging}) by an equality in
order to seek the most efficient safe merging possible and $x_{i-1}(t_i^m)$ is known (an explicit function of time).

Let $\bm\psi_{i}:=(\psi_{i,1},\psi_{i,2})^T$, $\bm\nu_{i}:=(\nu_{i,1},\nu
_{i,2})^T$ and define the costate $\bm\lambda_{i}:=(\lambda_{i}^{x},\lambda
_{i}^{v})^T$. The
Hamiltonian with the constraints adjoined is
\begin{equation}
\begin{aligned} H_i(\bm x_i,\bm\lambda_i, u_i) = &\frac{1}{2}u_i^2 + \lambda_i^xv_i + \lambda_i^vu_i \\&\!+\! \mu_i^a(u_i\! -\! u_{max})\! +\! \mu_i^b(u_{min}\! -\! u_i) \\&+ \mu_i^c(v_i - v_{max}) + \mu_i^d(v_{min} - v_i) \\&+ \mu_i^e(x_i + \varphi v_i -x_{i_p}) + \beta \end{aligned}
\end{equation}
The Lagrange multipliers $\mu_{i}^{a},\mu_{i}^{b},\mu_{i}^{c},\mu_{i}^{d},\mu_{i}^{e}$ are positive when the constraints are active and become 0 when the constraints are strict. Note that when the safety constraint (\ref{Safety}) becomes active, the expression above involves $x_{i_p}(t)$ in the last term. When $i=1$, the optimal trajectory is obtained without this term, since (\ref{Safety}) is inactive over all $[t_1^0,t_1^m]$. Thus, once the solution for $i=1$ is obtained (based on the analysis that follows), $x_1^{\ast}$ is a given function of time and available to $i=2$. Based on this information, the optimal trajectory of $i=2$ is obtained. Similarly, all subsequent optimal trajectories for $i >2$ can be recursively obtained based on $x_{i_p}^{\ast}(t)$ with $i_p < i-1$. As in Section A, we start with the case of
no active constraints, and then consider the effect of the safety constraint
(\ref{Safety}) becoming active.

\subsubsection{\textbf{Control, state, safety constraints not active}}

In this case, $\mu_{i}^{a}=\mu_{i}^{b}=\mu_{i}^{c}=\mu_{i}^{d}=\mu_{i}^{e}=0$.
Applying the optimality condition, we get the same results as
(\ref{OptimalUinLambda})-(\ref{Optimal_x}). 

Since $\psi_{i,2}$ is an explicit function of time ($x_{i-1}(t_i^m)$ is an explicit function of time), the transversality
condition is
\begin{equation}
\left.\bm\nu_i^T\frac{\partial \bm\psi_i}{\partial t} + H_{i}(\bm x_{i}(t),\bm\lambda_{i}(t),u_{i}(t))\right\vert
_{t=t_{i}^{m}}=0 \label{TransversalityInB}%
\end{equation}
with the costate boundary condition $\bm\lambda_{i}(t_{i}^{m})=[(\bm\nu
_{i}^T\frac{\partial\bm\psi_{i}}{\partial\bm x_{i}})^{T}]_{t=t_{i}^{m}}$.

We get $\frac{\partial \bm\psi_i}{\partial \bm x_i}$ and $\frac{\partial \bm\psi_i}{\partial t}$ by:
\begin{equation}
\frac{\partial \bm\psi_i}{\partial \bm x_i} = \left[
\begin{matrix}
1 & 0 \\
1 & \varphi
\end{matrix}
\right],
\frac{\partial \bm\psi_i}{\partial t} = \left[
\begin{matrix}
0 \\
-v_{i-1}(t_i^m)
\end{matrix}
\right]
\end{equation}

By the costate boundary condition, we have
\begin{equation}
\bm\lambda_i(t_i^m) = \left[
\begin{matrix}
\nu_{i,1}(t_i^m) + \nu_{i,2}(t_i^m) \\
\varphi \nu_{i,2}(t_i^m) 
\end{matrix}
\right]\label{Costate}%
\end{equation}

By (\ref{OptimalUinLambda})-(\ref{Optimal_x}), it follows that $\lambda_i^x(t_i^m) = a_i, \lambda_i^v(t_i^m) = -u_i(t_i^m)$, and by (\ref{Costate}), we have $\nu_{i,2}(t_i^m) = \frac{1}{\varphi}\lambda_i^v(t_i^m)$.

Then, the transversality
condition (\ref{TransversalityInB}) is explicitly rewritten as
\begin{equation}
\beta+a_{i}v_{i}(t_{i}^{m})-\frac{1}{2}u_{i}^{2}(t_{i}^{m})+\frac{1}{\varphi
}u_{i}(t_{i}^{m})v_{i-1}(t_{i}^{m})=0
\end{equation}

By Assumption 2, it follows that at $t=t_{i}^{m}$ we have $v_{i-1}(t_{i}%
^{m})=v_{i-1}(t_{i-1}^{m})$, a constant known to CAV $i$, and $x_{i-1}%
(t_{i}^{m})=v_{i-1}(t_{i-1}^{m})(t_{i}^{m}-t_{i-1}^{m})$ with $t_{i-1}^{m}$
also known to CAV $i$. Then, for each $i\in S(t)$, we need to solve the
following algebraic equations for $a_{i},b_{i},c_{i},d_{i}$ and
$t_{i}^{m}$:
\begin{equation}
\begin{aligned} &\frac{1}{2}a_i\cdot(t_i^0)^2 + b_it_i^0 + c_i = v_i^0,\\ &\frac{1}{6}a_i\cdot(t_i^0)^3 + \frac{1}{2}b_i\cdot(t_i^0)^2 + c_it_i^0+d_i = 0,\\ &\frac{1}{6}a_i\cdot(t_i^m)^3 + \frac{1}{2}b_i\cdot(t_i^m)^2 + c_it_i^m+d_i = L,\\ &v_{i-1}(t_{i-1}^m)(t_i^m\! -\! t_{i-1}^m) \!=\! \varphi(\frac{1}{2}a_i\cdot(t_i^m)^2 \!+\! b_it_i^m \!+\! c_i) \!+\!\delta\\ &\beta + 0.5a_i^2\cdot(t_i^m)^2 + a_ib_it_i^m + a_ic_i - 0.5(a_it_i^m + b_i)^2\\ &\qquad\qquad \qquad\qquad + \frac{1}{\varphi}(a_it_i^m + b_i)v_{i-1}(t_{i-1}^m) = 0. \end{aligned}\label{OptimalSolInB}%
\end{equation}

Observe that in this case there is no safety constraint involving CAVs $i$ and
$i-1$ for all $t\in\lbrack t_{i}^{0},t_{i}^{m})$ because they are in different
lanes and only the safe merging constraint is of concern. 

\textbf{Numerical Example:} We have also conducted simulations in MATLAB to study the solution of (\ref{OptimalSolInB}). The simulation parameters are $t_i^0 = 1s, v_i^0 = 20m/s, v_{i-1}(t_{i-1}^m) = 30m/s, t_{i-1}^m = 15s, \varphi = 1.8, \delta = 0, L = 400m, \beta = 2.667$ ($\alpha = 0.2573$). Similarly as (\ref{OptimalSolInA}), we can still get four, six or eight solutions depending on these parameters. There is also only one feasible solution, i.e., $t_i^m = 16.6856s$.

\begin{figure}[thpb]
	\centering
	\includegraphics[scale=0.5]{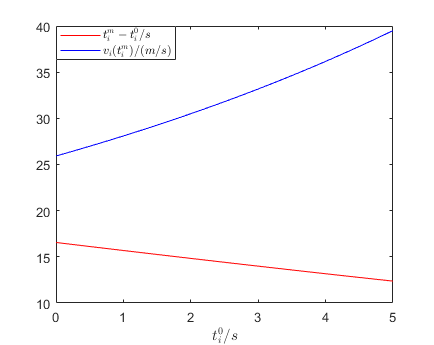}
	\caption{Optimal solutions for $t_i^0$ variation ($i-1 > i_p$).}
	\label{Bt0F}
\end{figure}

\begin{figure}[thpb]
	\centering
	\includegraphics[scale=0.5]{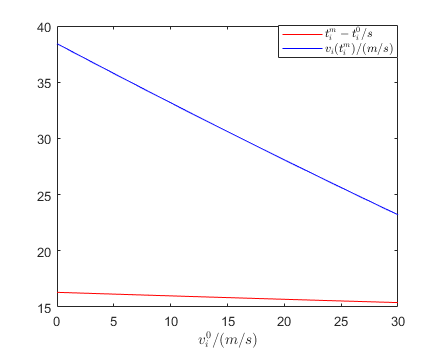}
	\caption{Optimal solutions for $v_i^0$ variation ($i-1 > i_p$).}
	\label{Bv0F}
\end{figure}

\begin{figure}[thpb]
	\centering
	\includegraphics[scale=0.4]{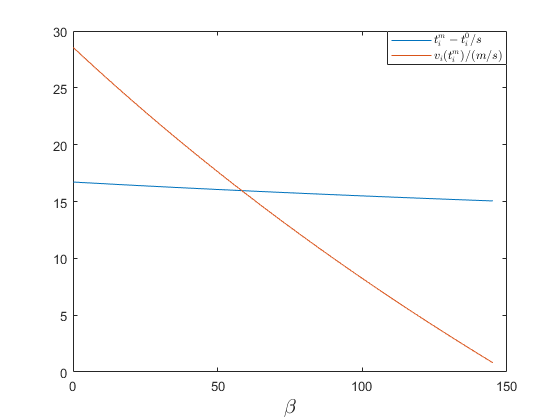}
	\caption{Optimal solutions for $\beta$ variation ($i-1 > i_p$).}
	\label{BbetaF}
\end{figure}

In this case, $t_i^0, v_i^0$ and $\beta$ will all affect the solutions. The simulation for the variation of $t_i^0$ is shown in Fig.\ref{Bt0F} ($v_i^0 = 20m/s, v_{i-1}(t_{i-1}^m) = 30m/s, t_{i-1}^m = 15s, \varphi = 1.8, L = 400m, \beta = 2.667 (\alpha = 0.2573)$), the simulation for the variation of $v_i^0$ is shown in Fig.\ref{Bv0F} ($t_i^0 = 1s, v_{i-1}(t_{i-1}^m) = 30m/s, t_{i-1}^m = 15s, \varphi = 1.8, L = 400m, \beta = 2.667 (\alpha = 0.2573)$), and the simulation for the variation of $\beta$ is shown in Fig.\ref{BbetaF} ($t_i^0 = 1s, v_i^0 = 20m/s, v_{i-1}(t_{i-1}^m) = 30m/s, t_{i-1}^m = 15s, \varphi = 1.8, L = 400m$).


We notice from Fig.\ref{Bv0F} and Fig.\ref{BbetaF} that the variation of $v_i^0$ and $\beta$ have few influence on the travel time $t_i^m - t_i^0$, which is due to the safe merging constraint (\ref{SafeMerging}).

If we want the speed of the vehicle $i$ to be equal to the speed of the vehicle $i-1$, i.e., $v_i(t_i^m) = v_{i-1}(t_{i-1}^m)$, we can either put constraint on $t_i^0$ or $v_i^0$. For example, we can make $v_i(t_i^m) = v_{i-1}(t_{i-1}^m)$ be a new constraint and take $v_i^0$ as a variable, then we can solve (\ref{OptimalSolInB}) together with this new constraint. In the simulation, $t_i^0 = 1s, v_{i-1}(t_{i-1}^m) = 30m/s, t_{i-1}^m = 15s, \varphi = 1.8, \delta = 0, L = 400m, \beta = 2.667 (\alpha = 0.2573)$. After solving these six nonlinear equations ,we can get $v_i^0 = 16.2005m/s, v_i(t_i^m) = 30m/s, t_i^m - t_i^0 = 15.8s$. We can also check for the state constraint and control constraint with the solutions.

Following from Theorem 1, we have the following theorem for $i-1$ and $i$:

\vspace{2ex} $\textbf{Theorem 4:}$ Under Assumptions 1-2, if CAVs $i$ and $i-1$ satisfy $v_{i}^{0}\leq
v_{i-1}^{0}$ and $t_{i}^{0}-t_{i-1}^{0}\geq\varphi+\frac{\delta}{v_{i}^{0}}$,
then, under optimal control (\ref{Optimal_u}) for both CAVs, the safe merging
constraint (\ref{SafeMerging}) is satisfied. \vspace{2ex}

\textbf{Remark 3:} Theorem 4 is useful for the case that the vehicle 
$i$ arrives much later than $i-1$, i.e.,
$t_{i}^{0}>>t_{i-1}^{0}$. In this case, if we still use the
optimal control solved by (\ref{OptimalSolInB}), the constraint (\ref{VehicleConstraints}) will most probably be violated. If Theorem 4 does not apply, we can also apply (\ref{Optimal_u}) for $i$ and check whether the safe merging constraint (\ref{SafeMerging}) will be satisfied or not. If yes, then we are done; otherwise, we can use the optimal control solved by (\ref{OptimalSolInB}).

\subsubsection{\textbf{Safety Constraint Active}}

Suppose that the safety constraint between CAVs $i$ and $i_{p}$ becomes active
at time $t_{1}\in(t_{i}^{0},t_{i}^{m})$ (where $t_{1}$ will be optimally
determined), i.e., $g_{i}(t_{1})=0$ with $g_{i}(t_{1})$ defined in
(\ref{GFunction}). As in section A, we can obtain the same optimal solutions
as (\ref{NewOptimal_u})-(\ref{NewOptimal_v}) for $t\geq t_{1}$ and the same optimal solutions as (\ref{OptimalSolBeforeT1}) for $t\in\lbrack t_{i}^{0}%
,t_{1})$, and Theorem 3 still holds.

In this case, we can always find an exit time $t_{2}$ from the safety
constrained arc on an optimal trajectory because this safe merging constraint between
$i$ and $i-1$ should be satisfied at $t_{i}^{m}$. Starting from
$t_{1}$, we can apply the optimal control derived from (\ref{OptimalSolInB})
but with different intial conditions. As in Sec.\ref{sec:constrained_A}, the safety constraint may be immediately violated, so we can obtain $t_{2}$ by solving
\begin{equation}
\begin{aligned} &\frac{1}{2}a_it_2^2 + b_it_2 + c_i = v_i^*(t_2),\\ &\frac{1}{6}a_it_2^3 + \frac{1}{2}b_it_2^2 + c_it_2+d_i = x_i^*(t_2),\\ &\frac{1}{6}a_i\cdot(t_i^m)^3 + \frac{1}{2}b_i\cdot(t_i^m)^2 + c_it_i^m+d_i = L,\\ &v_{i-1}(t_{i-1}^m)(t_i^m\! -\! t_{i-1}^m) \!=\! \varphi(\frac{1}{2}a_i\cdot(t_i^m)^2 \!+\! b_it_i^m \!+\! c_i) \!+\!\delta\\ &\beta + 0.5a_i^2\cdot(t_i^m)^2 + a_ib_it_i^m + a_ic_i - 0.5(a_it_i^m + b_i)^2\\ &\qquad\qquad \qquad\qquad + \frac{1}{\varphi}(a_it_i^m + b_i)v_{i-1}(t_{i-1}^m) = 0,\\ &a_it_2 + b_i = u_i^*(t_2). \end{aligned}\label{SolveForT2InB}%
\end{equation}
with $t_{1}$ to be optimally determined, where $x_{i}^{\ast}(t_{2}),v_{i}^{\ast}(t_{2})$ and
$u_{i}^{\ast}(t_{2})$ are optimal solutions from (\ref{NewOptimal_v}). $t_{2}$ is a function of $t_{1}$ when solving
(\ref{SolveForT2InB}). 

We can still apply Theorem 2 to find the infeasible interval $I_i$ for $i$ to exclude these $t_1$ that do not make $t_1$ be the first time that the safety constraint (\ref{Safety}) becomes active.

Therefore, we can apply (\ref{OptimalSolBeforeT1}) for $t\in[t_{i}^{0},t_{1}%
)$, apply (\ref{NewOptimal_u}) for $t\in[t_{1},t_{2})$ and apply the optimal
control solved by (\ref{SolveForT2InB}) for $t\in[t_{2},t_{i}^{m}]$. Then we
can get the optimal solutions for $J_{i}^{*}(t_{1})$, {\small
\begin{equation}\label{eqn:J*_B}
\begin{aligned} J_i^*(t_1) = \beta(t_i^m - t_i^0) + \int_{t_0}^{t_1}\frac{1}{2}(u_i^*(t))^2 dt \\+ \int_{t_1}^{t_2}\frac{1}{2}(u_i^*(t))^2 dt + \int_{t_2}^{t_i^m}\frac{1}{2}(u_i^*(t))^2 dt \end{aligned}
\end{equation}
}
The optimal solution for $t_{1}$ is obtained by finding $t_{1}$ that
minimizes $J_{i}^{\ast}(t_{1})$. By (\ref{SolveForT2InB}), it follows that $t_{i}^{m}$
is dependent on $t_{1}$.

We can also summarize the method of finding the optimal $t_1^*$, $t_2^*$, $x_{i}^*(t), v_{i}^*(t)$ and $u_{i}^*(t)$ by the following algorithm:

\begin{algorithm}[t]
	\caption{Safety constrained optimal trajectory, $i_p > i-1$} 
	\hspace*{0.02in} {\bf Input:} 
	$t_i^0, v_i^0, x_{i_p}^*(t), v_{i_p}^*(t),  u_{i_p}^*(t)$,\\ $x_{i-1}^*(t), v_{i-1}^*(t),  u_{i-1}^*(t)$\\
	\hspace*{0.02in} {\bf Output:} 
	$t_1^*, x_{i}^*(t), v_{i}^*(t),  u_{i}^*(t)$, $t_2^*$
	\begin{algorithmic}[1]
		\State solve (\ref{OptimalSol})
		\If{CAV $i_p$ is under unconstrained optimal control (\ref{Optimal_u})} 		
		\State $I_i:=\{t_1\in (t_i^0,t_i^m] |u_i(t_1) + \varphi a_i > u_{i_p}^*(t_1)\}$
		\Else
		\State $I_i = \{t_1\in (t_i^0,t_i^m]|\exists t\in[t_i^0,t_1), g_i(t) > 0\}$
		\EndIf
		\State get feasible set $F_i:=(t_i^0,t_i^m)\setminus I_i$ for $t_1$
		\State solve (\ref{NewOptimalUVX}) and (\ref{SolveForT2InB})
		\State get $J_i^*(t_1)$ by (\ref{eqn:J*_B})
		\State solve for $t_1^*$ over $F_i$ (and $t_2^*$)
		\State result = $t_1^*, x_{i}^*(t), v_{i}^*(t),  u_{i}^*(t) , t_2^*$
		\State \Return result
	\end{algorithmic}
\end{algorithm}

\textbf{Numerical Example:} There three vehicles $i\in S(t)$, $i_p(\ne i-1) \in S(t)$ and $i-1\in S(t)$ with parameters $t_{i_p}^0 = 0s$, $v_{i_p}^0 = 20m/s,$ $t_{i-1}^0 = 0.1s, v_{i-1}^0 = 20m/s$, $t_i^0 = 2.55s, v_i^0 = 28m/s$ let $\varphi = 1.8s, ,\delta = 0m, L = 400m$, and $\beta = 2.667$ ($\alpha = 0.2573$). The vehicle $i_p$ and $i$ are in the same lane, the vehicle $i-1$ is in the different lane with respect to $i_p$ and $i$. Therefore, $i_p = i-2$ in the FIFO queue.

\begin{figure}[thpb]
	\centering
	\includegraphics[scale=0.5]{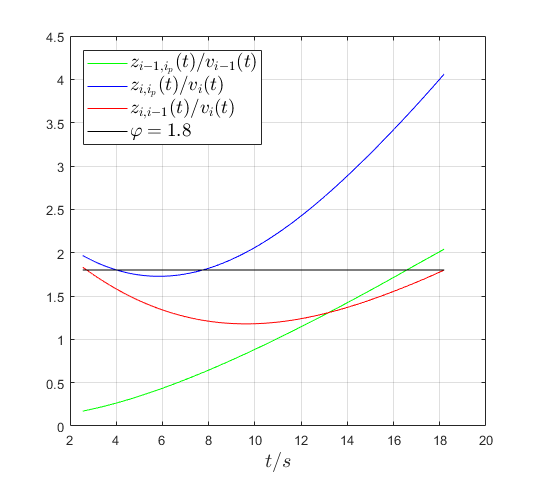}
	\caption{The safety constraint (safe merging) profile for vehicles $i,i-1,i_p$ if $i$ is under unconstrained optimal control solved by (\ref{OptimalSolInB}).}
	\label{MergingF}
\end{figure}

If we apply the optimal controller for $i_p$ solved by (\ref{OptimalSolInA}), the optimal controller for $i-1$ and $i$ solved by (\ref{OptimalSolInB}), then we can get their safety constraint and safe merging profile, as shown in Fig.\ref{MergingF}.

\begin{figure}[thpb]
	\centering
	\includegraphics[scale=0.5]{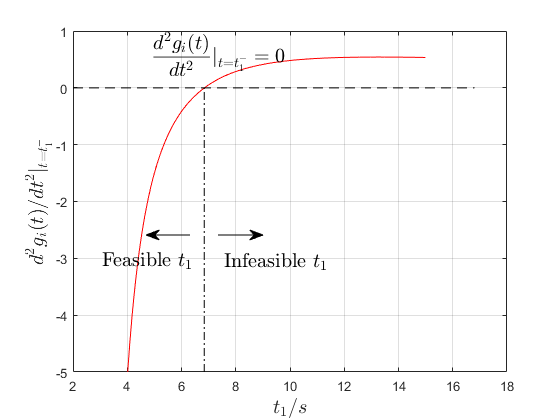}
	\caption{The second order derivative of $g_i(t)$ with respect to $t$ at $t = t_1^-$. }
	\label{fig:ddG_B}%
\end{figure}

The safe merging constraint between $i-1$ and $i_p$, $i-1$ and $i$ should only be satisfied at the merging point, as shown in Fig.\ref{MergingF} (green and red lines). However, the safety constraint between $i_p$ and $i$ should always be satisfied. We notice the safety constraint between $i_p$ and $i$ is violated for some time, as shown in the second frame of Fig.\ref{MergingF} (blue line). Therefore, we need to solve the optimal solution again. We use (\ref{OptimalSolBeforeT1}) for $t\in[t_i^0,t_1)$, (\ref{NewOptimalUVX}) for $t\in[t_1,t_2)$ and the optimal solution by (\ref{SolveForT2InB}) for $t\in[t_2,t_i^m]$. Firstly, we check whether there is infeasible interval $I_i$ for $t_1$, as shown in Fig.\ref{fig:ddG_B}.

\begin{figure}[thpb]
	\centering
	\includegraphics[scale=0.5]{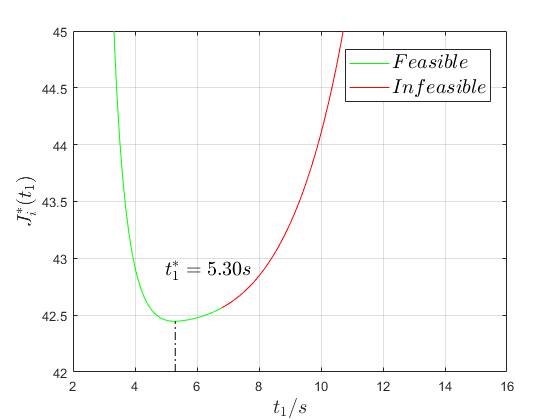}
	\caption{ $J_i^*(t_1)$ with respect to $t_1$. }
	\label{fig:J_B}%
\end{figure}

It follows from Fig.\ref{fig:ddG_B} that the infeasible interval $I_i = (6.84, t_i^m]$ does exist in this case. The optimal objective function with respect to $t_1$ is shown in Fig.\ref{fig:J_B} and we get $t_1^{\ast} = 5.30s$ following from  Fig.\ref{fig:J_B}.

\begin{figure}[thpb]
	\centering
	\includegraphics[scale=0.5]{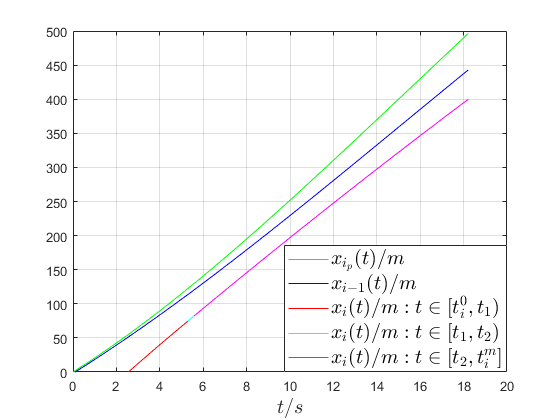}
	\caption{The position profiles for $i$, $i-1$ and $i_p$. }
	\label{fig:pos_B}%
\end{figure}

\begin{figure}[thpb]
	\centering
	\includegraphics[scale=0.5]{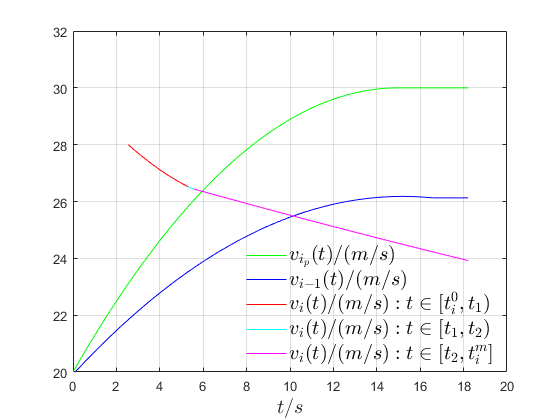}
	\caption{The speed profiles for $i$, $i-1$ and $i_p$. }
	\label{fig:speed_B}%
\end{figure}

\begin{figure}[thpb]
	\centering
	\includegraphics[scale=0.5]{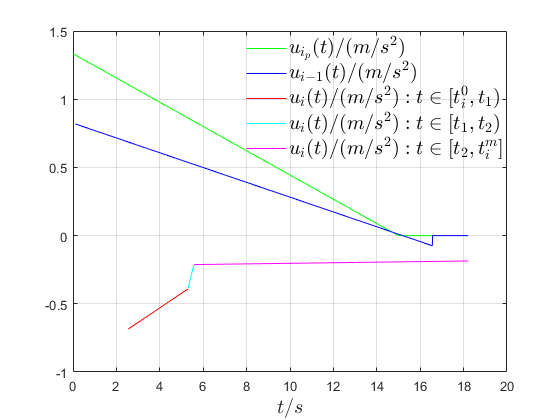}
	\caption{The control profiles for $i$, $i-1$ and $i_p$. }
	\label{fig:control_B}%
\end{figure}

\begin{figure}[thpb]
	\centering
	\includegraphics[scale=0.5]{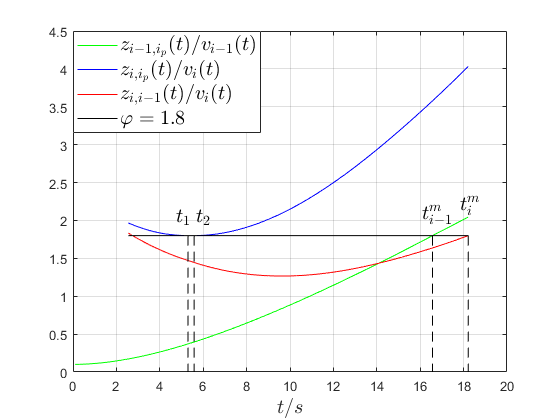}
	\caption{The safety and safe merging profiles for $i$, $i-1$ and $i_p$. }
	\label{fig:safety_B}%
\end{figure}

We continue to study the state and safety constraint profiles at $t_1^{\ast} =  5.30s$ and $t_2^{\ast} =  5.5794s$, as shown in Fig.\ref{fig:pos_B}-\ref{fig:safety_B}. 


\section{SIMULATION EXAMPLES}

We have used the Vissim microscopic multi-model traffic flow simulation tool
as a baseline to compare with the optimal control approach we have developed.
The car following model in Vissim is based on \cite{Wiedemann1974} and
simulates human psycho-physiological driving behavior.

The simulation parameters used are as follows: $L=400m,$ $\varphi=1.8,$
$\delta=0,$ $v_{max}=30m/s,$ $v_{min}=10m/s,$ $u_{max}=3.924m/s^{2}$ and
$u_{min}=-3.924m/s^{2}$. The simulation under optimal control is conducted in
MATLAB by using the same vehicle input and initial conditions as in Vissim.
The CAVs arrive randomly with 600 CAVs per hour arrival rate for both lanes.

The simulation results regarding the performance under optimal control
compared to that in Vissim  are summarized in Table I. We can see that the objective
function defined in (\ref{CAV_problem}) is significantly improved under
optimal control compared to the Vissim simulation for both cases
($\alpha=0.26$ and $\alpha=0.41$). The same applies to the average travel times.

\begin{table}[ptb]
\caption{Objective function comparison}
\par
\begin{center}%
\begin{tabular}
[c]{|c||c||c||c||c|}\hline
Items & \multicolumn{2}{|c||}{OC}  & \multicolumn{2}{|c|}{Vissim}\\\hline
Weight& $\alpha$=0.26&$\alpha$=0.41  &$\alpha$=0.26 &$\alpha$=0.41 \\\hline\hline
Ave. time/s & 17.0901 & 15.2297 &\multicolumn{2}{|c|}{ 30.9451}\\\hline
Main time/s & 17.1304 & 15.2609 & \multicolumn{2}{|c|}{23.7826}\\\hline
Merg. time/s & 17.0489 & 15.1978 & \multicolumn{2}{|c|}{38.2667}\\\hline
Ave. $\frac{1}{2}u_{i}^{2}(t)$ & 5.6979 & 11.9167 & \multicolumn{2}{|c|}{20.0918}\\\hline
Main $\frac{1}{2}u_{i}^{2}(t)$ & 5.8349 & 12.3077 & \multicolumn{2}{|c|}{9.4066}\\\hline
Merg. $\frac{1}{2}u_{i}^{2}(t)$ & 5.5580 & 11.5171 & \multicolumn{2}{|c|}{31.0144}\\\hline
Ave. obj. & 38.4308 & 55.1110 & 76.8200&109.5478\\\hline
Main obj. & 38.6219 & 55.4402 & 54.5736&80.6316\\\hline
Merg. obj.& 38.2448 & 54.7745 & 99.5606&139.1065\\\hline
\end{tabular}
\end{center}
\par
\end{table}

Recognizing that $1/2u^{2}(t)$ is only an approximation of the actual fuel
consumption of a vehicle, we have used the polynomial metamodel proposed in
\cite{Kamal2013} for a more accurate evaluation of fuel consumption as a
function of both $v_{i}(t)$ and acceleration $u_{i}(t)$. This model is defined
as
\begin{equation}
\dot{f}_{v}(t)=\dot{f}_{cruise}(t)+\dot{f}_{accel}(t)\label{FuelModel}%
\end{equation}
where
\[
\dot{f}_{cruise}(t)=\omega_{0}+\omega_{1}v_{i}(t)+\omega_{2}v_{i}%
^{2}(t)+\omega_{3}v_{i}^{3}(t)
\]%
\[
\dot{f}_{accel}(t)=(r_{0}+r_{1}v_{i}(t)+r_{2}v_{i}^{2}(t))u_{i}(t)
\]
and $\omega_{0}$, $\omega_{1}$, $\omega_{2}$, $\omega_{3}$, $r_{0}$, $r_{1}$
and $r_{2}$ are positive coefficients (we used the values reported in
\cite{Kamal2013}). It is assumed that during braking from a high velocity when  $u_i(t) < 0$, no fuel is consumed. The comparison results are shown in Table II. As is to be
expected, fuel consumption under optimal control is larger compared to that
obtained in the Vissim simulation, since the form used for the objective
function in (\ref{CAV_problem}) is different from (\ref{FuelModel}). It
remains unclear what an accurate fuel consumption model is and this is the
subject of ongoing and future work aiming at appropriate modifications of
(\ref{CAV_problem}).

\begin{table}[ptb]
\caption{Fuel consumption comparison}
\par
\begin{center}%
\begin{tabular}
[c]{|c||c||c||c|}\hline
Items & OC ($\alpha$=0.26) & OC ($\alpha$=0.41) & Vissim\\\hline\hline
Ave. fuel/mL & 48.6124 & 68.3194 & 36.9954\\\hline
Main fuel/mL & 48.0726 & 67.2866 & 42.6925\\\hline
Merg. fuel/mL & 49.1642 & 69.3752 & 31.1717\\\hline
\end{tabular}
\end{center}
\par
\end{table}

\section{CONCLUSIONS}

We have derived a decentralized optimal control solution for the traffic
merging problem that jointly minimizes the travel time and energy consumption
of each CAV and guarantees that a speed-dependent safety constraint is always
satisfied. Under certain simple-to-check condition in Theorems 1,4, we have
shown that the safety constraint remains inactive and computation is
simplified. Otherwise, we have still derived a complete solution that may
include one or more arcs where the safety constraint is active. We have not
taken into account speed and acceleration constraints for each CAV, which will
be incorporated in future work by including appropriate arcs in the optimal
trajectory as in \cite{Zhang2016}. Ongoing research is exploring the use of
approximate solutions (e.g., the use of control barrier functions) as an
alternative to an optimal control solution if the latter becomes
computationally burdensome or if the use of more complex objective functions
or more elaborate vehicle dynamics makes an optimal control approach
prohibitive.  Lastly, we will investigate the case where only a fraction of the
traffic consists of CAVs, similar to the study in \cite{Zhang2018CCTA}.

\addtolength{\textheight}{-12cm}





\bibliographystyle{plain}
\bibliography{Hamilton}

\begin{thebibliography}{10}

\bibitem{Bryson1969}
Bryson and Ho.
\newblock {\em Applied Optimal Control}.
\newblock Ginn Blaisdell, Waltham, MA, 1969.

\bibitem{Cao2015}
W.~Cao, M.~Mukai, and T.~Kawabe.
\newblock Cooperative vehicle path generation during merging using model
  predictive control with real-time optimization.
\newblock {\em Control Engineering Practice}, 34:98--105, 2015.

\bibitem{Chen2017}
W.~Chen, Z.~Zhao, Z.~liu, and Peter C.~Y. Chen.
\newblock A novel assistive on-ramp merging control system for dense traffic
  management.
\newblock In {\em Proc. IEEE Conference on Industrial Electronics and
  Applications}, pp. 386--390, Siem Reap, 2017.

\bibitem{Kamal2013}
M.~Kamal, M.~Mukai, J.~Murata, and T.~Kawabe.
\newblock Model predictive control of vehicles on urban roads for improved fuel
  economy.
\newblock {\em IEEE Transactions on Control Systems Technology},
  21(3):831--841, 2013.

\bibitem{Levine1966}
W.~Levine and M.~Athans.
\newblock On the optimal error regulation of a string of moving vehicles.
\newblock {\em IEEE Transactions on Automatic Control}, 11(13):355--361, 1966.

\bibitem{Malikopoulos2018}
A.~A. Malikopoulos, C.~G. Cassandras, and Yue~J. Zhang.
\newblock A decentralized energy-optimal control framework for connected and
  automated vehicles at signal-free intersections.
\newblock {\em Automatica}, 2018(93):244--256, 2018.

\bibitem{Milanes2012}
V.~Milanes, J.~Godoy, J.~Villagra, and J.~Perez.
\newblock Automated on-ramp merging system for congested traffic situations.
\newblock {\em IEEE Transactions on Intelligent Transportation Systems},
  12(2):500--508, 2012.

\bibitem{Mukai2017}
M.~Mukai, H.~Natori, and M.~Fujita.
\newblock Model predictive control with a mixed integer programming for merging
  path generation on motor way.
\newblock In {\em Proc. IEEE Conference on Control Technology and
  Applications}, pp. 2214--2219, Mauna Lani, 2017.

\bibitem{Ntousakis2016}
I.~A. Ntousakis, I.~K. Nikolos, and M.~Papageorgiou.
\newblock Optimal vehicle trajectory planning in the context of cooperative
  merging on highways.
\newblock {\em Transportation Research}, 71, Part C:464--488, 2016.

\bibitem{Raravi2007}
G.~Raravi, V.~Shingde, K.~Ramamritham, and J.~Bharadia.
\newblock {\em Merge algorithms for intelligent vehicles. In: Sampath, P.,
  Ramesh, S. (Eds.), Next Generation Design and Verification Methodologies for
  Distributed Embedded Control Systems}.
\newblock Springer, Waltham, MA, 2007.

\bibitem{Rathgeber2015}
C.~Rathgeber, F.~Winkler, X.~Kang, and S.~Muller.
\newblock Optimal trajectories for highly automated driving.
\newblock {\em International Journal of Mechanical, Aerospace, Industrial,
  Mechatronic and Manufacturing Engineering}, 9(6):946--952, 2015.

\bibitem{Torres2015}
J.~Rios-Torres, A.A. Malikopoulos, and P.~Pisu.
\newblock Online optimal control of connected vehicles for efficient traffic
  flow at merging roads.
\newblock In {\em Proc. IEEE 18th International Conference on Intelligent
  Transportation Systems}, pp. 2432--2437, Las Palmas, Spain, 2015.

\bibitem{Scarinci2014}
R.~Scarinci and B.~Heydecker.
\newblock Control concepts for facilitating motorway on-ramp merging using
  intelligent vehicles.
\newblock {\em Transport Reviews}, 34(6):775--797, 2014.

\bibitem{Schrank2015}
B.~Schrank, B.~Eisele, T.~Lomax, and J.~Bak.
\newblock The 2015 urban mobility scorecard.
\newblock Texas A\&M Transportation Institute, 2015.

\bibitem{Tideman2007}
M.~Tideman, M.C. van~der Voort, B.~van Arem, and F.~Tillema.
\newblock A review of lateral driver support systems.
\newblock In {\em Proc. IEEE Intelligent Transportation Systems Conference},
  pp. 992--999, Seatle, 2007.

\bibitem{Varaiya1993}
P.~Varaiya.
\newblock Smart cars on smart roads: problems of control.
\newblock {\em IEEE Transactions on Automatic Control}, 38(2):195--207, 1993.

\bibitem{Vogel2003}
K.~Vogel.
\newblock A comparison of headway and time to collision as safety indicators.
\newblock {\em Accident Analysis \& Prevention}, 35(3):427--433, 2003.

\bibitem{Waard2009}
D.~De Waard, C.~Dijksterhuis, and K.~A. Broohuis.
\newblock Merging into heavy motorway traffic by young and elderly drivers.
\newblock {\em Accident Analysis and Prevention}, 41(3):588--597, 2009.

\bibitem{Weng2016}
J.~Weng, S.~Xue, and X.~Yan.
\newblock Modeling vehicle merging behavior in work zone merging areas during
  the merging inplementation period.
\newblock {\em IEEE Transactions on Intelligent Transportation Systems},
  17(4):917--925, 2016.

\bibitem{Wiedemann1974}
R.~Wiedemann.
\newblock Simulation des straßenverkehrsflusses.
\newblock In {\em Proc. of the Schriftenreihe des tnstituts fir Verkehrswesen
  der Universitiit Karlsruhe (In German language)}, 1974.

\bibitem{Zang2009}
X.~Zang.
\newblock The short-term traffic volume forecasting for urban interchange based
  on rbf artificial neural networks.
\newblock In {\em Proc. IEEE Conference on Mechatronics and Automation}, pp.
  2607--2611, Changchun, 2009.

\bibitem{Zhang2018}
Yue~J. Zhang and C.~G. Cassandras.
\newblock A decentralized optimal control framework for connected automated
  vehicles at urban intersections with dynamic resequencing.
\newblock In {\em Proc. 57th IEEE Conference on Decision and Control}, 2018.
\newblock To appear.

\bibitem{Zhang2018CCTA}
Yue~J. Zhang and C.~G. Cassandras.
\newblock The penetration effect of connected automated vehicles in urban
  traffic: an energy impact study.
\newblock In {\em Proc. 2018 IEEE Conference on Control Technology and
  Applications}, pp. 620--625, Copenhagen, Denmark, 2018.

\bibitem{Zhang2017}
Yue~J. Zhang, C.~G. Cassandras, and A.~A. Malikopoulos.
\newblock Optimal control of connected and automated vehicles at urban traffic
  intersections: A feasibility enforcement analysis.
\newblock In {\em Proc. of the American Control Conference}, pp. 3548--3553,
  Seattle, 2017.

\bibitem{Zhang2016}
Yue~J. Zhang, A.~A. Malikopoulos, and C.~G. Cassandras.
\newblock Optimal control and coordination of connected and automated vehicles
  at urban traffic intersections.
\newblock In {\em Proc. of the American Control Conference}, pp. 6227--6232,
  Boston, 2016.

\end{thebibliography}

\end{document}